\documentclass[sn-mathphys,pdflatex]{sn-jnl}% Math and Physical Sciences Reference Style
\jyear{2021}%

\theoremstyle{thmstyleone}%
\theoremstyle{thmstyletwo}%

\theoremstyle{thmstylethree}%

\usepackage{subcaption} %Para generar subfigure
\usepackage[normalem]{ulem} % aux: para tachar \sout

\begin{document}

\title[Semi-collapse of turbulent fountains..]{Semi-collapse  of turbulent fountains in stratified media and the mechanisms to control their dynamics}

%\author[1]{\fnm{Luis G.} \sur{Sarasua}}\email{sarasua@fisica.edu.uy}
\author*[1]{\fnm{Luis G.} \sur{Sarasúa}}\email{sarasua@fisica.edu.uy}

%\author[1]{\fnm{Daniel} \sur{Freire Caporale}}
\author*[1]{\fnm{Daniel} \sur{Freire Caporale}}\email{dfreire@fisica.edu.uy}

%\author[2]{\fnm{Nicasio} \sur{Barrere}}
\author[2]{\fnm{Nicasio} \sur{Barrere}}\email{nbarrere@fisica.edu.uy}

%\author[1]{\fnm{Arturo C.} \sur{Marti}}
\author*[1]{\fnm{Arturo C.} \sur{Martí}}\email{marti@fisica.edu.uy}

\affil*[1]{\orgdiv{Instituto de Física}, \orgname{Universidad de la República}, \country{Uruguay}}
\affil*[2]{\orgdiv{Centro Universitario del Este}, \orgname{Universidad de la República}, \country{Uruguay}}

\abstract{Turbulent fountains are widespread natural phenomena with
  numerous industrial applications. Extensive research has focused on
  the temporal evolution and maximum height of these fountains, as
  well as their dependence on Reynolds and Froude numbers. However,
  the minimum height of the surrounding ambient fluid, which is
  removed by the fountain due to the entrainment effect, has received
  little attention. In this study, we investigate the dependence of
  this minimum height on the characteristics of the fountain and
  demonstrate how to control it. Our findings present important
  implications for technological applications of turbulent fountains,
  particularly in contaminant withdrawal. We discuss the potential of
  our results to improve the efficiency of such applications.  }

\keywords{turbulent fountain, stratified media, semi-collapse, fluid control, contaminant withdrawal}

\maketitle

\section{Introduction} \label{sec:intro}

A fountain is a vertical buoyant jet in which the buoyancy force and
the jet’s initial velocity act in opposite directions. On the other
hand, the flow is a plume if the buoyancy force acts in the same
direction as the jet velocity.  Fountains and plumes are frequently
encountered in nature and technical applications.  Since the fluid
dynamics in stratified media presents problems of considerable
interest across several fields, turbulent fountains and plumes in
stratified media have been the research subject for
decades~\cite{Woods, Kaye, Burridge, Turner, Morton1956, Morton1959,
  Morton1973, Bloomfield1998, Bloomfield1999, Richards, Camassa,
  Carroll, Ezhova}.  In particular, the behaviour of the pollutants
ejected into the atmosphere is of great interest due to the effects
these emissions produce on human health and nature in general
\cite{Vallero}. Although air pollution is caused by emissions from
different kind of sources, fountain flows often appear in the process
of contaminant ejection as a consequence of the use of stacks
\cite{Vallero}.  The evolution of the contaminants in the atmosphere
is usually described by models like the Gaussian plume model, which
depends on parameters that are determined by empirical
relations. Although these models are proven to be useful, these do not
allow the study of effects of variations in the characteristics of the
flow at the source.
% Hunt2001, Hunt2005,Lin

Fountain dynamics in stratified media can be outlined as follows. At
an initial stage, the fountain decelerates due to the opposing
buoyancy force and the entrainment of ambient fluid reaching a maximum
height at which the vertical momentum is zero. Then the flow reverses
direction and falls as an annular plume around the fountain core.
Depending on the initial fluxes of momentum and buoyancy and the
initial stratification profile, the fountain spreads outwards at a
non-zero spreading height, above the source level, or the flow
collapses, i.e., it falls to the source or ground level (the terrain).

Some aspects of the turbulent fountains in the quasi-steady regime can
be described using the well-known MTT model proposed by Morton, Taylor
and Turner \cite{Morton1956, Morton1959, Morton1973}, which describe
the evolution of volume, momentum, and buoyancy fluxes in
fountains. In this model, it is assumed that the horizontal velocity
at which the ambient fluid enters the fountain is proportional to the
vertical velocity in the fountain, with a proportionality coefficient
%$\alpha$
called the entrainment coefficient. Although successful in predicting
the evolution in a uniform ambient and the maximum height in
plumes~\cite{Kaye}, the MTT model do not describe the dynamics after
the vertical velocity reverses its direction. Bloomfield and Kerr
~\cite{Bloomfield1998} proposed that the spreading height can be
obtained by matching it to the height where the fluid density of the
environment is equal to the fluid density at the maximum height. This
condition is used to estimate $h_{m}$ and $h_{sp}$ combining different
models. This approach may be considered a first-order because it does
not take into account the mixing between fountain and ambient fluid in
the downflow that occurs after the fountain reverses its direction.

Some years later, Kaminski et al.~\cite{Kaminski} developed an
expression for the entrainment coefficient depending on three
parameters that can be experimentally determined. A comparison between
the predictions based on this expression and the experimental data was
given in \cite {Carazzo} for the case of a homogeneous medium. Mehaddi
et al. \cite{Mehaddi} studied fountains in stratified environments and
obtained expression for the maximum height. However, the spreading
behaviour of the fountain was not considered in this
investigation. Papanicolau et al. \cite{papanicolaou2010spreading}
conducted an experimental study on the collapse and spreading of
turbulent fountains and performed a comparison with those obtained in
\cite{Bloomfield1998}. As various authors have pointed out
\cite{Turner1966, Telford, Reeuwijk}, to assume a constant entrainment
coefficient is an approximation due to its dependence on the
turbulence intensity, and, as a consequence, it can vary with the rise
of the fountain. Recently, Sarasua et al. \cite{Sarasua} proposed a
model that generalizes the model of Morton et al. to determine the
dependence of the maximum and the spreading height with the parameters
involved. This model determines the critical conditions for the
collapse of the fountain, i.e., when the jet falls to the source
level, using a parameter that measures the mixing of the jet with the
environment along the downflow.

Despite these contributions, a study of the dependence of the minimal
height of the spreading flow, here referred to as \textit{critical
  height} with the parameters controlling the flow, is not available
yet. For configurations where the collapse regime does not occur, the
critical height determines whether the fluid removed by the
entrainment mechanism and later deviated to the spreading cloud will
be again in contact with the ground or not.

This study focuses on a detailed analysis of the flow structure of
turbulent fountains, specifically, the maximum, spreading, and
critical heights, using fully validated numerical simulations. We
investigate several configurations of turbulence levels and introduce
a novel parameter, the dimensionless \textit{lightness}, which depends
on the temperature difference between the impinging fountain and the
ambient fluid at the inlet. Our findings are abridged in a diagram
that results a powerful tool for the design of technological
applications of turbulent fountains in stratified media. Based on this
diagram, we propose strategies to modify the fountain conditions and
control its subsequent dynamics, particularly, the final regime
developed. Our work presents significant implications for practical
applications of turbulent fountains in the industry.

The structure of this work is as follows. First, in
Section~\ref{sec:description}, we provide a detailed description of
our case study. In Section~\ref{subseq:dimensiones_problemas}, we
analyse the characteristics of the domain, the experimental setup and
procedure, the implemented numerical simulations, and their full
validation with the experimental results. Subsequently, in
Section~\ref{sec:results}, we present and analyse the results
obtained. Specifically, we introduce a dimensionless parameter, the
``lightness'', report on the height variation with the relevant
parameters, and summarise the results in a diagram aimed at
controlling the characteristic fountain dynamics heights. Finally, in
Section~\ref{sec:conc}, we draw our conclusions.

\section{Problem description} \label{sec:description}

In engineering applications involving pollutant emissions, it is of
importance to warranty that the generated spreading flux does not fall
below certain level. Therefore, in addition to the maximum and
spreading heights, $h_m$ and $h_{sp}$, respectively, we defined a
third characteristic height, the
%\textit{minimal height}
\textit{critical height}, $h_c$ ($h_c \leq h_{sp}$), as the minimum
height that the spreading cloud reaches, which develops after the flow
reversion falling and stabilisation at $h_{sp}$. For instance, if $h_c
\sim 0$
%(see Figs.~\ref{fig:scheme_collapse} and \ref{fig:scheme_semi_collapse})
a potential hazard will occur
%(Fig.~\ref{fig:scheme_semi_collapse})
since  fluid ejected falls back to the ground.

\begin{figure}[htb!]
        \centering
        \begin{subfigure}[b]{0.32\textwidth}
     \includegraphics[width=\linewidth]{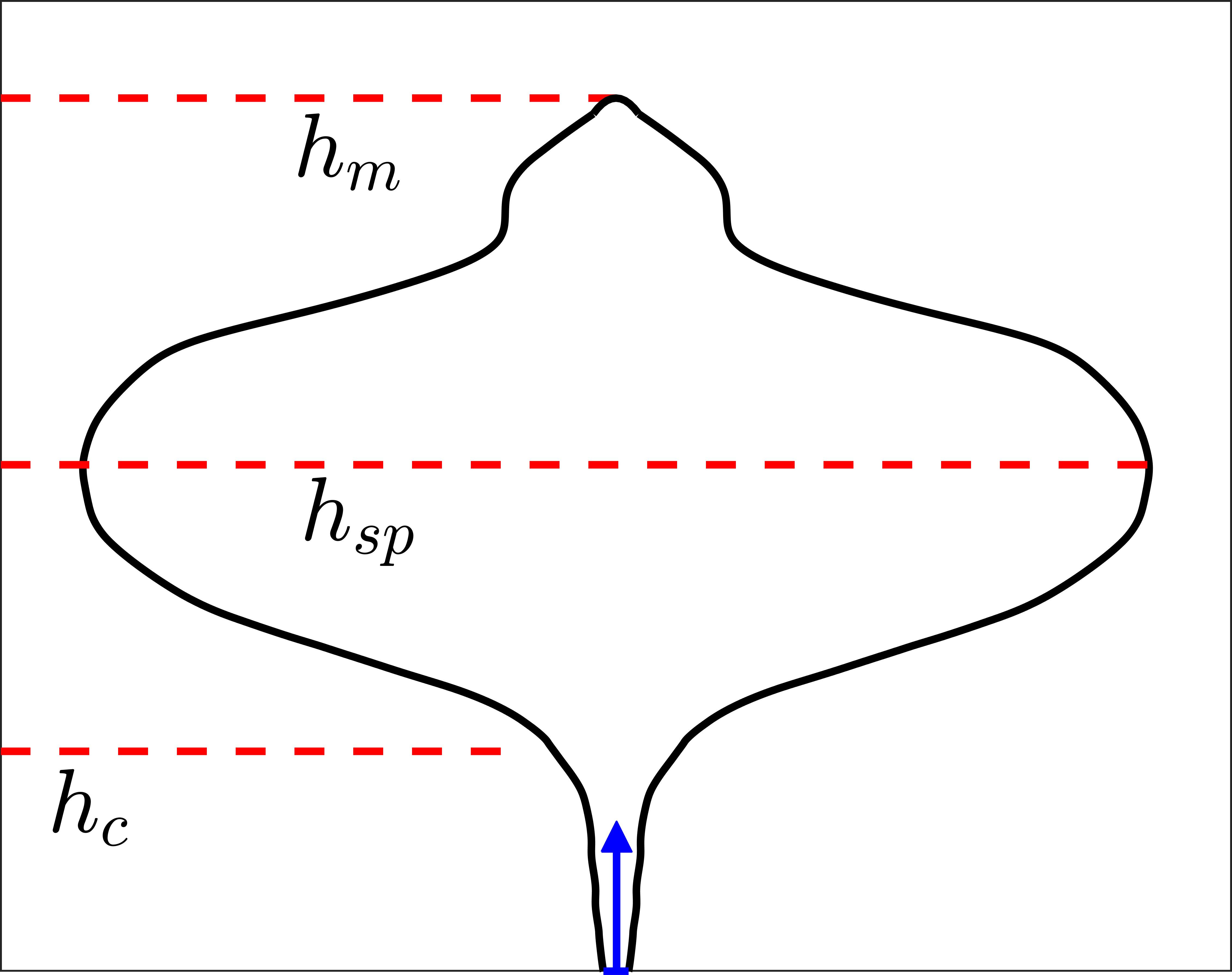}
     %\includegraphics[width=\linewidth]{figures/adim/contorno_2D_alturas_noCollapse_sinEjes_V3.png}
    %\caption{No collapse.}
    \caption{}
		\label{fig:scheme_no_collapse}
	\end{subfigure}
        \hfill 
	\begin{subfigure}[b]{0.32\textwidth}
  \includegraphics[width=\linewidth]{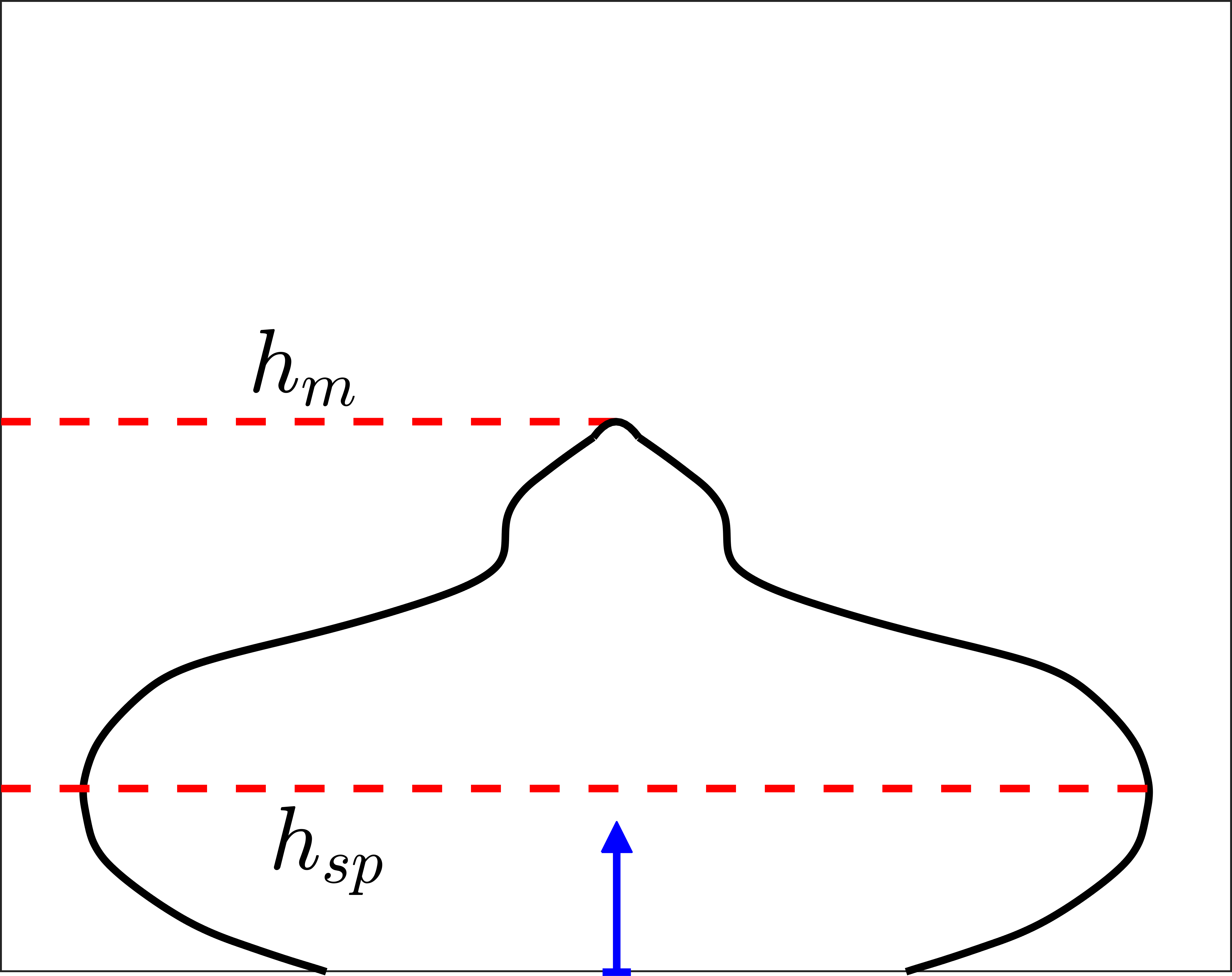}
		%\caption{Semi-collapse.}
        \caption{}
		\label{fig:scheme_semi_collapse}
	\end{subfigure}
	\hfill
	\begin{subfigure}[b]{0.32\textwidth}
  \includegraphics[width=\linewidth]{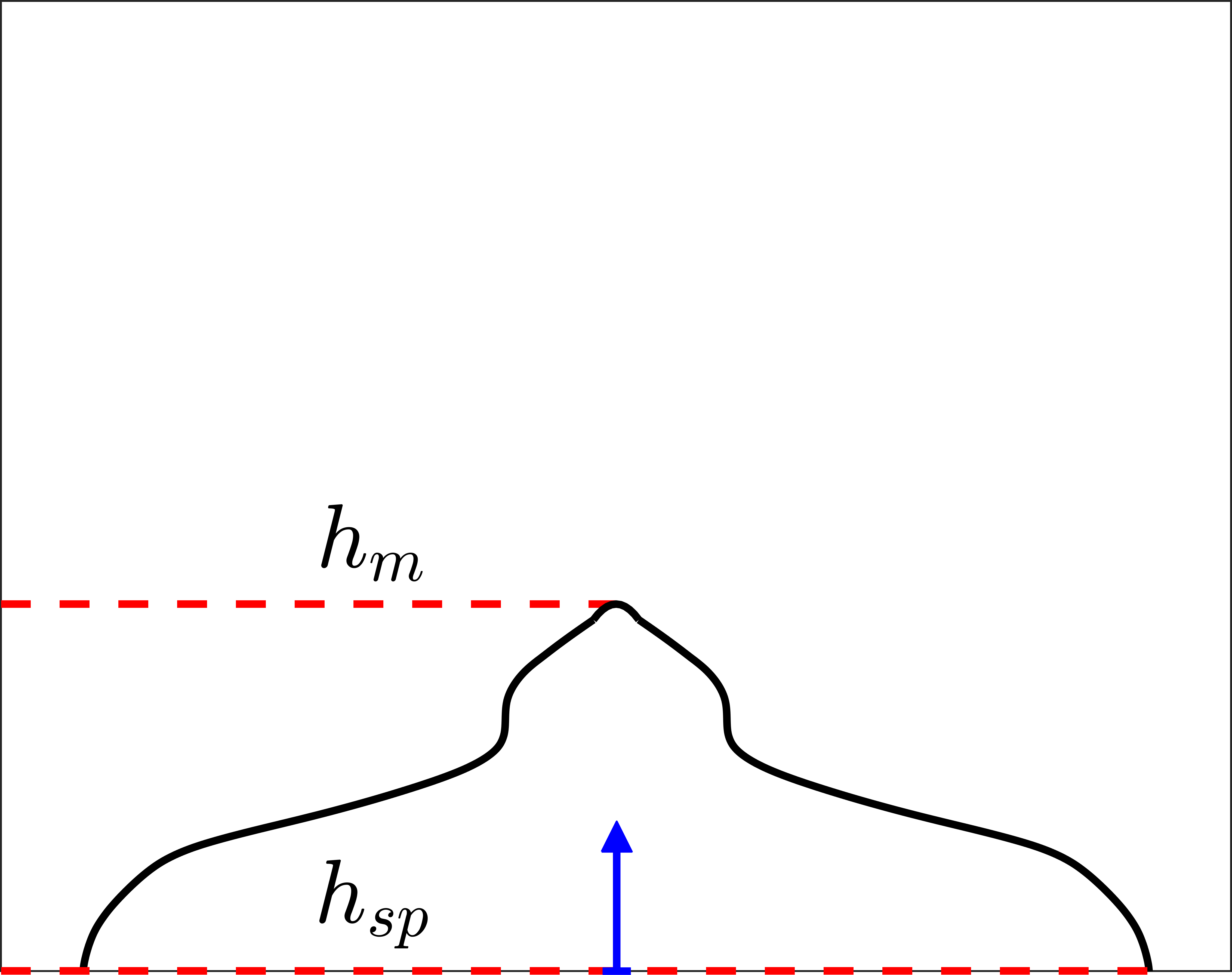}
		%\caption{Collapse.}
        \caption{}
		\label{fig:scheme_collapse}
	\end{subfigure}
	\caption{The three possible dynamics of the developed fountain
          at developed stages:
 %given a passive tracer concentration tolerance:
    %$\phi_{tol}$
    (a) no collapse when $h_{sp}>0$ and $h_c>0$; (b) semi-collapse
          when $h_{sp}>0$ and $h_c=0$; (c) collapse when $h_{sp}=0$
          (and so it does $h_c=0$).  Black lines correspond to the
          fountain contour
    %$\mathscr{C}$
    and the upward-pointing blue arrows indicate the fountain inflow
    (the gravity points downwards).}
        \label{fig:schemes}
\end{figure}

Three different regimes, sketched in Fig.~\ref{fig:schemes}, can be
developed by the fountain. First, there is the case in which the
extracted fluid does not come into contact with the ground surface
again, so-called the non-collapse regime, discriminated by $h_ {sp}>
0$ and $h_ {c}> 0$ (Fig.~\ref{fig:scheme_no_collapse}). Second, the
semi-collapse regime, Fig.~\ref{fig:scheme_semi_collapse}, in which,
although the front of the spreading flow does not make contact with
the ground ($ h_ {sp}> 0 $), its lower edge does ($h_c = 0$), that is,
the ejected flow comes back into contact with the ground. Finally, the
third case, Fig.~\ref{fig:scheme_collapse} is the collapse regime, in
which $h_{sp}=0$ and $h_{c}=0$.

%-------------------------------------
%-------------------------------------

%
For engineering applications devoted to the removal of contaminated
air from the vicinity of the ground (lower ambient strata) it is
crucial to elucidate the fountain's conditions at the entrance that
guarantee the non-collapse regime.  In the present work, we aimed at
finding the subsequent flow regime based on the flow conditions at the
inlet. For this purpose, we performed numerical simulations,
previously validated with experiments, where the temperature profile
of the environment stratification was kept constant and each
simulation case was performed for a given value of $T_{in}$ and the
turbulence level.

\section{Problem configuration} \label{subseq:dimensiones_problemas}

The experimental setup was designed based on scaling a technological
application called Selective Inverted Sink (SIS)\cite{sis}, which
represents a remarkable breakthrough utilising turbulent fountains to
mitigate frost damage in agriculture under radiation frost conditions.

This section is organized as follows. First, we present detailed
experimental measurements conducted in the laboratory in
Section~\ref{sec:experimentos}. Secondly, we explain the computational
simulations performed in Section~\ref{sec:simulaciones}. Finally, in
Section~\ref{sec:validacion_simulaciones}, we compare the experimental
and numerical results to validate the latter calculations.
\subsection{Experimental Setup} \label{sec:experimentos}

We present laboratory experiments carried out in a rectangular acrylic
tank of dimensions $40$~cm $\times$ $40$~cm $\times$ $50$~cm in width
($x$) $\times$ depth ($y$) $\times$ height) ($z$) filled with
water. The temperature of the ambient fluid was arranged to have a
linear profile, with $T_{cold}=15$~$^{\circ}$C at the lower boundary
($z=0$) and $T_{hot}=27.5$~$^{\circ}$C at the upper boundary
($z=50$~cm), leading to a temperature gradient of
$25$~$^{\circ}$C$\cdot$m$^{-1}$. Hence, the initial temperature
profile of the quiescent ambient fluid was given by $T(t=0,z)=
\partial_zT \cdot z + T_{cold}$.

\begin{figure}[htb!]
	\centering
        \includegraphics[width=0.55\linewidth,keepaspectratio]{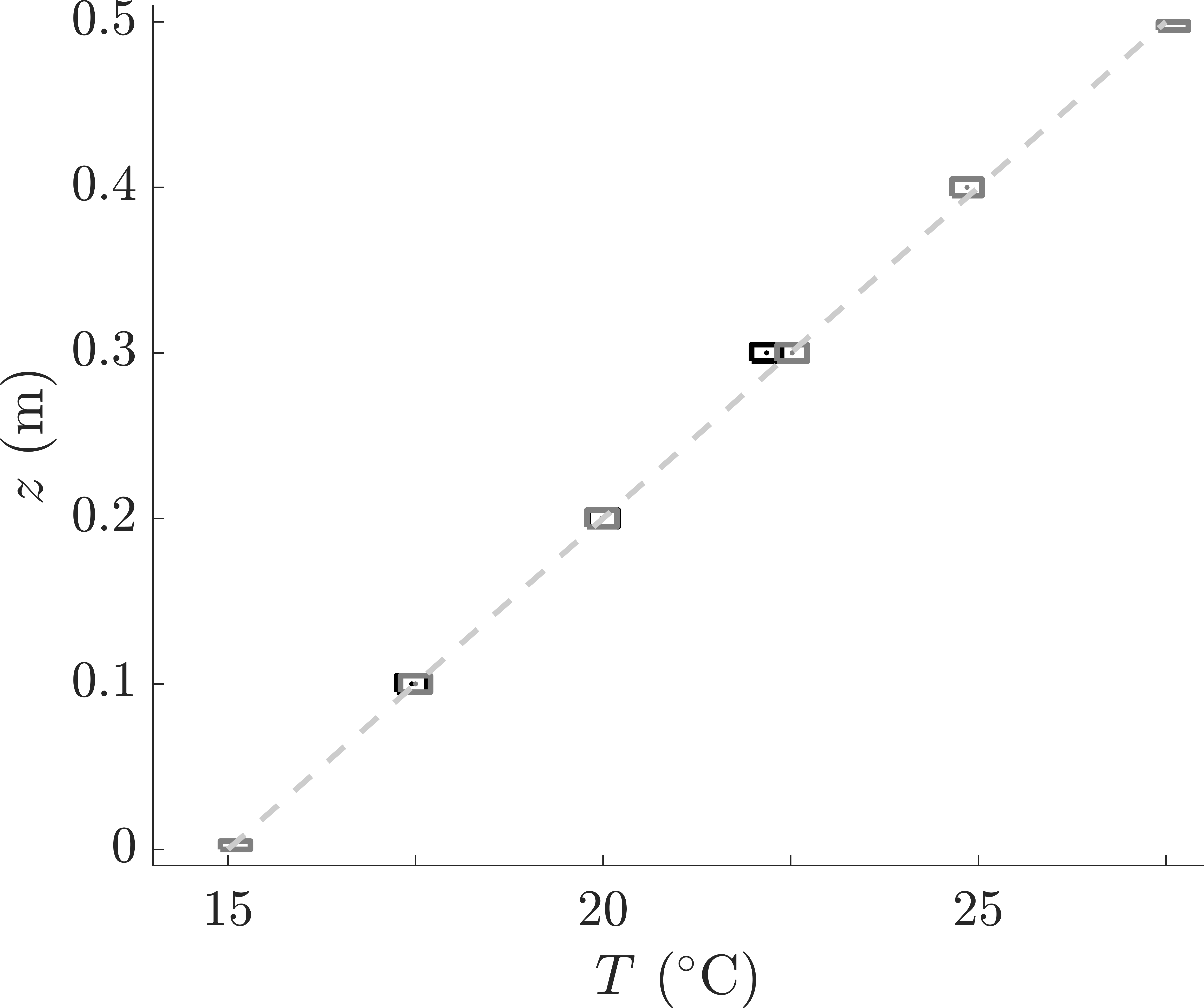}
	%\caption{Experimental temperature stratification profile prior to injecting the fountain at an inlet temperature of $T_{in}=15$~$^{\circ}$C. The rectangles denote the experimental measurements with their corresponding error bars, while the dashed line represents the linear fit of the data.}
	\caption{Typical temperature stratification profile observed
          experimentally before injecting the fountain into the system
          at an inlet temperature of
          $T_{in}=15$$^{\circ}$C. Rectangles indicate experimental
          measurements with their corresponding uncertainties while
          the dashed line represents a linear fit. The two sets of
          rectangles (black and gray) correspond to two vertical
          container edges where thermocouples were installed every
          10~cm.}
	\label{fig:T_profile_experiments}
\end{figure}

To establish a stable stratification, we followed a meticulous
procedure. Starting from the lowermost layers of water, we inserted a
$1$~kW electrical resistance and gradually heated the fluid while
monitoring its temperature using a thermocouple. Once the target
temperature was reached (depending on the vertical position of the
resistance), we carefully moved the resistance upwards to the next
fluid layer, repeating this procedure until we reached the highest
layer of fluid at $z=50$~cm. The bottom and top plates of the
container were controlled using cooler and heater, respectively, which
were regulated by relays. We insulated the tank using 10~cm deep
expanded polystyrene panels and monitored the temperature gradient
using thermocouples installed every 10~cm along two vertical edges of
the tank. By following this meticulous procedure, we successfully
established a quiescent ambient fluid with a stable linear vertical
stratification in temperature. Fig.~\ref{fig:T_profile_experiments}
shows a typical temperature profile obtained at the end of this
procedure.

%%%%%%%%%%%%%%%%%%%%%%%%%%%%%%%%%%%%%%%%%%%%

The fountain was generated by injecting water vertically at
$T_{in}=15$~$^{\circ}$C, at a stable flow rate of
$\dot{q}_{in}=5.5$~$\mathrm{cm}^3/s$, through a circular nozzle with a
diameter of $D=8$~mm, positioned at the center of the tank bottom
plate. Figures~\ref{fig:tinta_experimental_3figs}(a)--(c) displays
experimental images for the grid configuration obtained using the dye
tracer technique.

We investigated the flow using two methods: dye (tracer) visualization
and Digital Particle Image Velocimetry
(DPIV)~\cite{westerweel1997fundamentals,gui1998generating,adrian2011particle}. The
former method allowed us to track the evolution of the inflow (colored
fluid) as its dynamics evolved, and mixing and diffusion occurred. The
latter enabled us to determine the velocity fields of the seeded
fluids. To capture the velocity fields, we illuminated the fluid with
a 500~mW green LASER sheet of 2~mm thickness and captured images using
a CMOS camera at 4~fps. We processed the images using the open-source
package OpenPIV~\cite{ben2020openpiv}. We performed DPIV experiments
in both free and grid configurations. In the grid configuration, we
placed an 80-mesh screen with a wire diameter of 0.18~mm transversally
at the jet inlet to generate different degrees of
turbulence~\cite{freire2010effect}. To ensure the reliability of the
results, the experiments were conducted with 2-3 repetitions for each
configuration and technique. Figures~\ref{fig:campos_vel}~(a) and (c)
present an example of the experimentally measured horizontal and
vertical components, respectively, of the velocity field for the free
configuration at 60 s.

In this work, we present both experimental and numerical studies on
the flow of confined turbulent fountains in a stratified medium. To
extrapolate the results for studying the efficiency of the SIS device,
the measurement period was approximately 120 s, during which the
characteristic heights became stabilized, and no significant
recirculation velocities were observed in the vicinity of the
container walls.

%\textcolor{red}{En la Fig.~XXXX mostramos fotografías experimentales para el caso GRID? con visualización con trazador de color rojo.}

\subsection{Computer Simulations Setup} \label{sec:simulaciones}

%We performed extensive numerical experiments based on the open-source package caffa3d.MBRi~\cite{usera2008parallel,mendina2014general} which implements the Finite Volume Method~\cite{ferziger2020finite}. The domain dimensions, the flow rate at the inlet port as well as the properties of the ambient and the vertically injected fluids, were set as those from the experiments described in Sec.~\ref{sec:experimentos}.

\begin{figure}[htb!]
	\centering
	\begin{subfigure}[b]{0.390\textwidth}
		\centering
		\includegraphics[width=\linewidth]{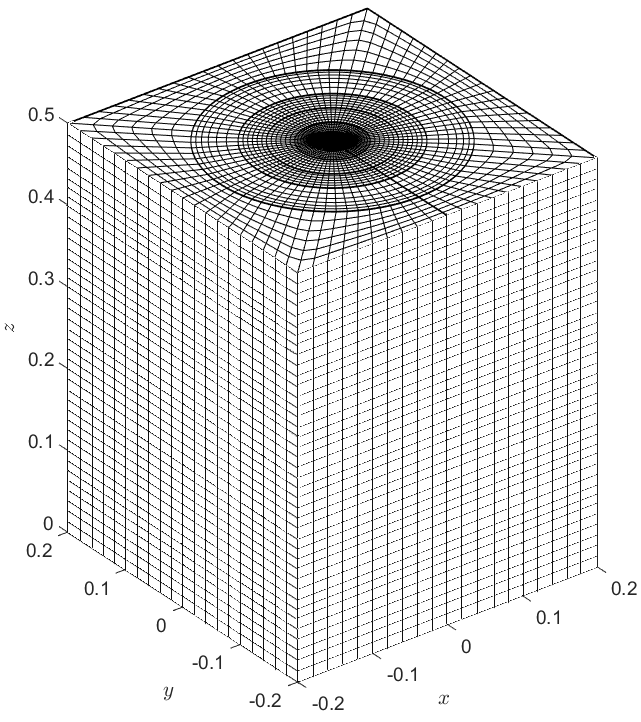}
		%\caption{View of $\phi$ over the $(x,z)$-plane.}
		\caption{}
%		\label{fig:constantforcesRA}
	\end{subfigure}
	%\hfill
	\hspace{1pc}
	\begin{subfigure}[b]{0.390\textwidth}
		\centering
		\includegraphics[width=\linewidth]{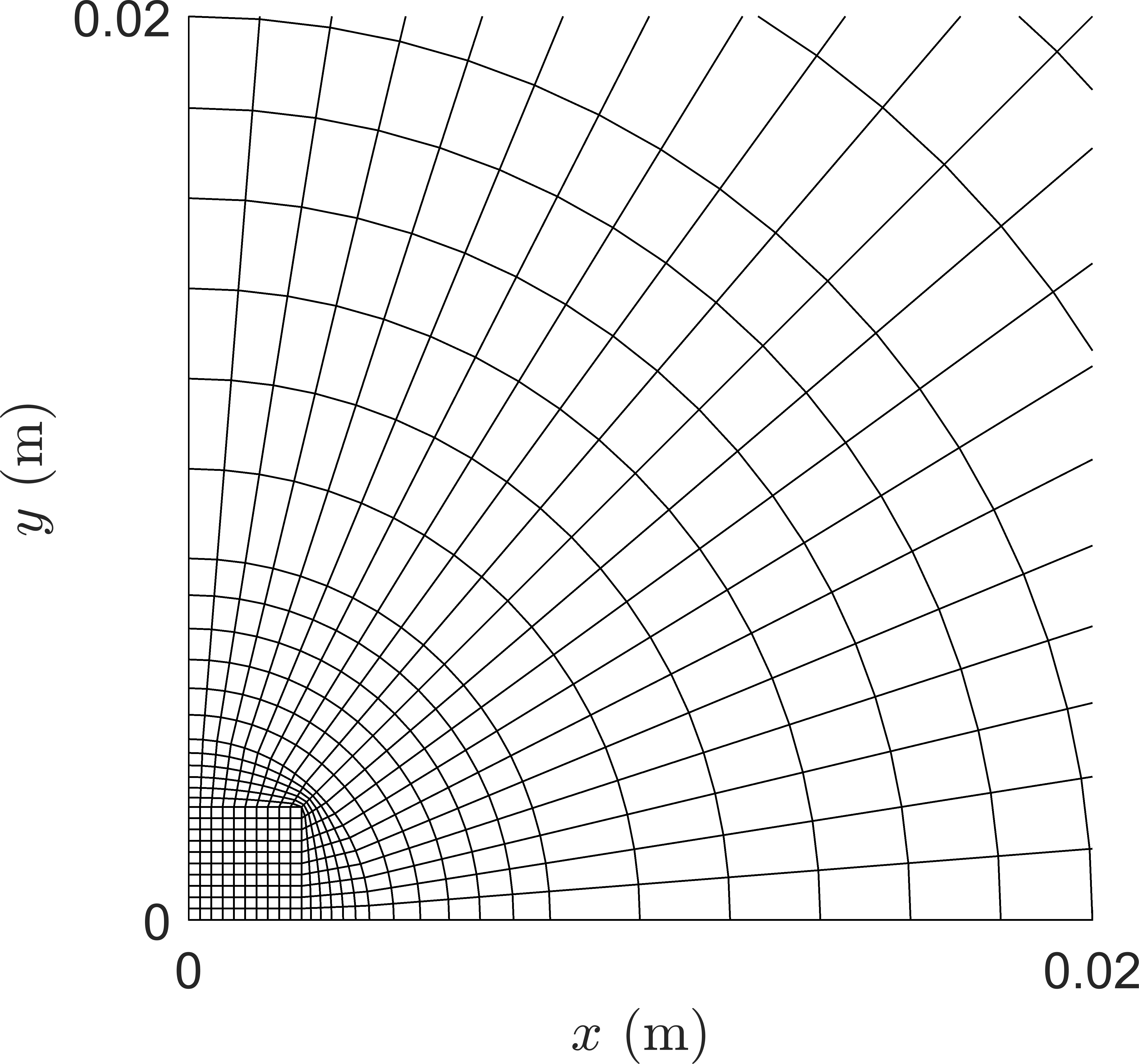}
		%\caption{View of $\phi$ over the $(x,y)$-plane.}
		\caption{}
%		\label{fig:sineforcesRA2}
	\end{subfigure}
	% \caption{Two views of the passive tracer field $\phi$ at time $t=100$~s for the configuration $(\xi=0,u^{\prime}/U=1\%)$. On the left (right), the view is from the $y$ ($z$) axis.}
	\caption{Distribution of domain-discretization cells shown from two different perspectives: (a) a perspective view and (b) a projection onto the $(x,y)$-plane.}
	\label{fig:malla_ejemplo}
\end{figure}

We conducted extensive numerical simulations using the open-source
package caffa3d.MBRi~\cite{usera2008parallel,mendina2014general} that
implements the Finite Volume Method~\cite{ferziger2020finite}. The
fluid properties in our simulations were set to those of pure water,
consistent with the experimental conditions. The domain dimensions and
the inlet flow rate in our simulations were set to match those of the
experiments described in Sec.~\ref{sec:experimentos}. The calculations
used a time step of 0.05 s, and the computational domain was
discretized as represented in Fig.~\ref{fig:malla_ejemplo} into
$8\times 10^6$ hexahedral cells, ensuring mesh independence by
utilizing both a coarser and a finer grid with similar results.

\begin{figure}[htb!]
	\centering
	\begin{subfigure}[b]{0.442\textwidth}
		\centering
		\includegraphics[width=\linewidth]{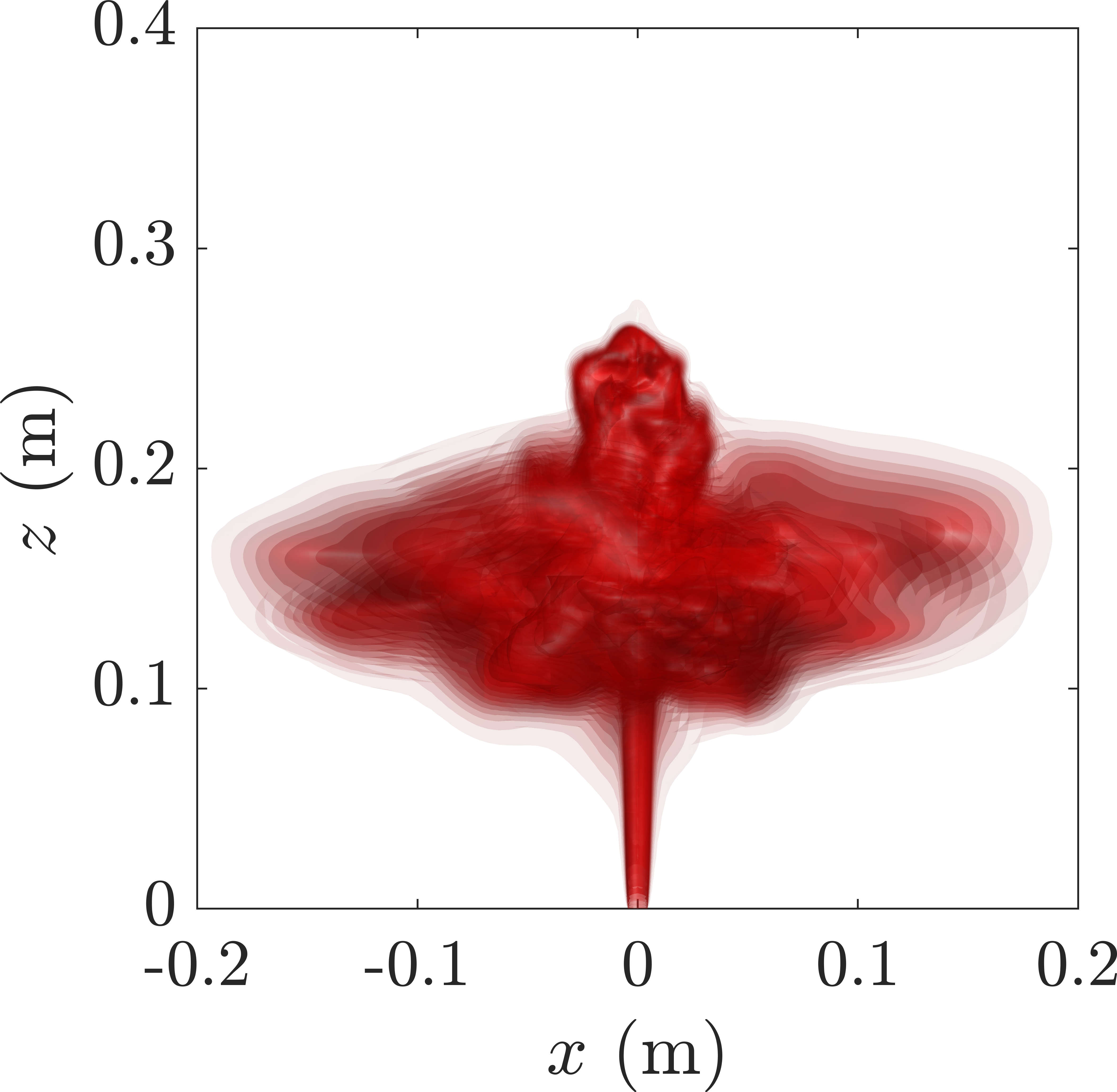}
		%\caption{View of $\phi$ over the $(x,z)$-plane.}
		\caption{}
	%	\label{fig:constantforcesRA}
	\end{subfigure}
	%\hfill
	\hspace{1pc}
	\begin{subfigure}[b]{0.442\textwidth}
		\centering
		\includegraphics[width=\linewidth]{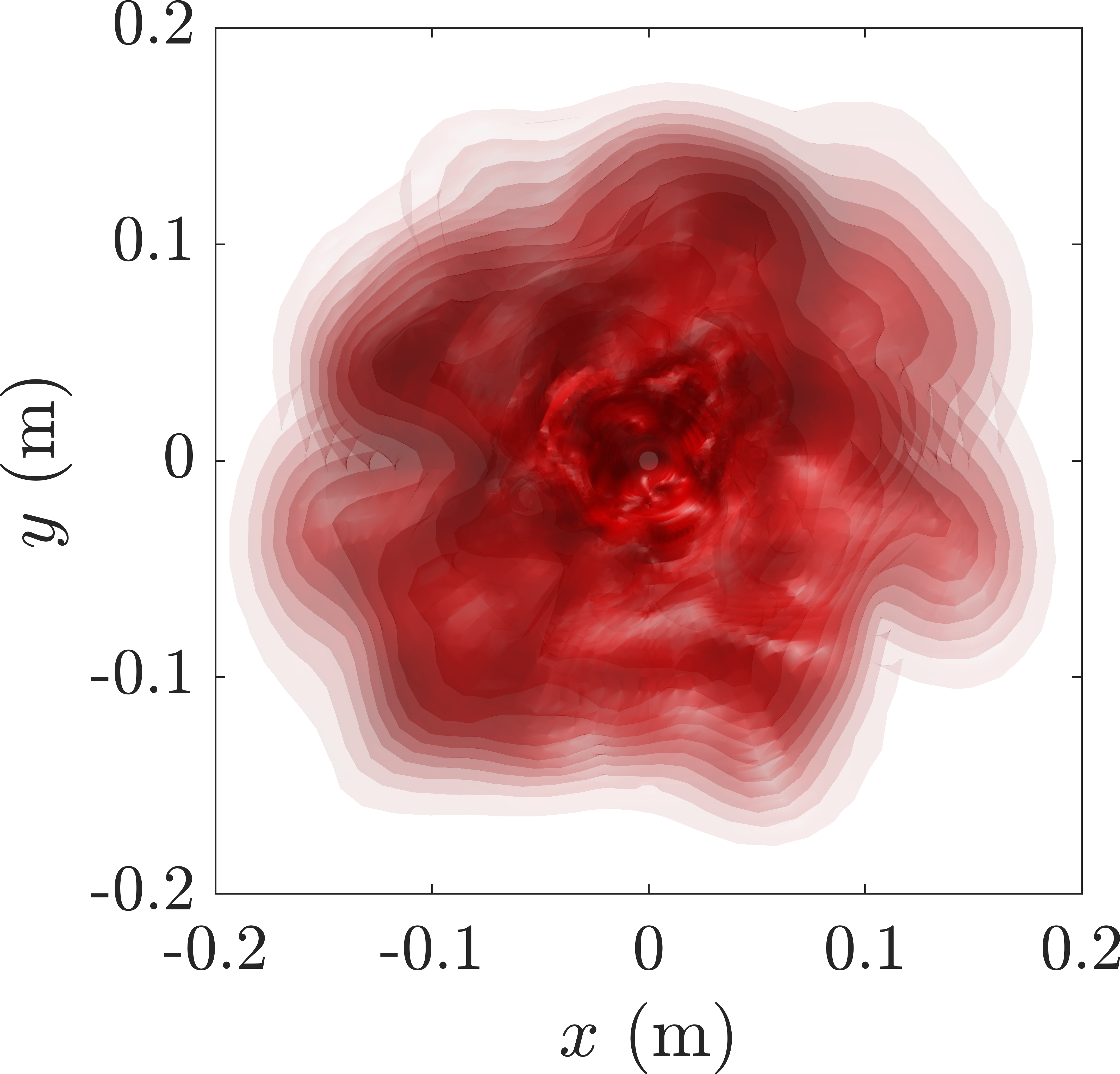}
		%\caption{View of $\phi$ over the $(x,y)$-plane.}
		\caption{}
		\label{fig:sineforcesRA2}
	\end{subfigure}
	% \caption{Two views of the passive tracer field $\phi$ at time $t=100$~s for the configuration $(\xi=0,u^{\prime}/U=1\%)$. On the left (right), the view is from the $y$ ($z$) axis.}
	\caption{Snapshots of the passive tracer field $\phi$:
		lateral,  $(x,z)$-plane, (a) and top perspective $(x,y)$-plane  (b).
		Parameter values: $t=100$~s, $\xi=0,u^{\prime}/U=1$.}
	%\caption{Two projections of the passive tracer field $\phi$. On the left (right), $\phi$ is projected onto the $(x,z)$-plane ($(x,y)$-plane).}
	\label{fig:3d_images_fountain}
\end{figure}

Turbulence was modeled using the Smagorinsky large-eddy
model~\cite{smagorinsky1963general}, where the constant factor in the
subgrid viscosity $\mu_\mathrm{sgs}$ was set to $C_S=0.16$.  In order
to replicate the behavior of the dye tracer observed in our
experiments, a passive scalar field represented by $\phi$ was employed
to indicate the concentration of dye in the fluid flow. Initially,
$\phi$ was assigned a value of 1 at the fountain inlet and 0
elsewhere. Subsequently, the interplay of advection, mixing, and
diffusion caused $\phi$ to vary between 0 and 1 over time within the
computational domain. A representative example of the computed flows
is depicted in Fig.~\ref{fig:3d_images_fountain}.

As explained in the previous section, the turbulence intensity of the
fountain at the inlet was controlled during the experiments by means
of a grid positioned at the entrance. For the numerical simulations,
this turbulence intensity was synthetically adjusted by introducing
random fluctuations, denoted as $\mathbf{u^{\prime}}=(u_x, u_y, u_z)$,
to the mean velocity at the inlet, i.e., $\mathbf{U_{in}}=\left<
\mathbf{U_{in}} \right>+\mathbf{u^{\prime}}$. At every simulation time
step, $u_x$, $u_y$, and $u_z$ were randomly assigned to each cell
located along the inlet port, with a uniform distribution.  For a
given turbulence level $u^{\prime}/U\in [0,1]$, $u_x$, $u_y$, and
$u_z$ are set randomly at each cell of the inlet port and at every
time step, so that $u_i/U \in [-u^{\prime}/U, +u^{\prime}/U]$, where
$\left\langle u_i \right\rangle=0$ for $i\in {x, \ y, \ z}$. We
employed a uniform distribution to generate such random fluctuations
in our calculations.  For a given turbulence level $u^{\prime}/U\in
[0,1]$, $u_x$, $u_y$, and $u_z$ are set randomly at each cell of the
inlet port and at every time step, so that $u_i/U \in [-u^{\prime}/U,
  +u^{\prime}/U]$, where $\left\langle u_i \right\rangle=0$ for $i\in
{x, \ y, \ z}$. We employed a uniform distribution to generate such
random fluctuations in our calculations.  Assuming a constant inlet
flow rate $\dot{q}_{in}$, we set the velocity at each cell within the
inlet as $\mathbf{U_{in}}=( u_x, u_y, U+u_z )$, where
$U=\frac{\dot{q}_{in}}{\pi D^2/4}$ is the average vertical velocity at
every cell within the inlet boundary condition. We conducted
simulations at turbulence levels of $u^{\prime}/U = 0$, $1$, $2$, $4$,
$10$, and $20\%$. By adjusting the turbulence intensity, we were able
to replicate the experimental results for the free and grid flow
configurations with turbulence intensity values of $u^{\prime}/U = 1$
and $20\%$, respectively. These results are shown in
Section~\ref{sec:validacion_simulaciones}.

\subsection{Validation of Computer Simulations} \label{sec:validacion_simulaciones}

As detailed in Sec.~\ref{sec:experimentos}, we studied two
configurations, the free and grid configurations, at $\xi=0$. We
employed two different measurement techniques, dye tracer and DPIV,
although not simultaneously, and repeated each experiment 2-3
times. Similarly, as explained in Sec.\ref{sec:simulaciones}, we
computed the passive tracer and velocity fields throughout the
numerical simulations. We validated the computer simulation results by
comparing them with the corresponding experimental results.

\begin{figure}[htb!]
	\centering
	\begin{subfigure}[b]{0.305\textwidth}
		\centering
		\includegraphics[width=\linewidth]{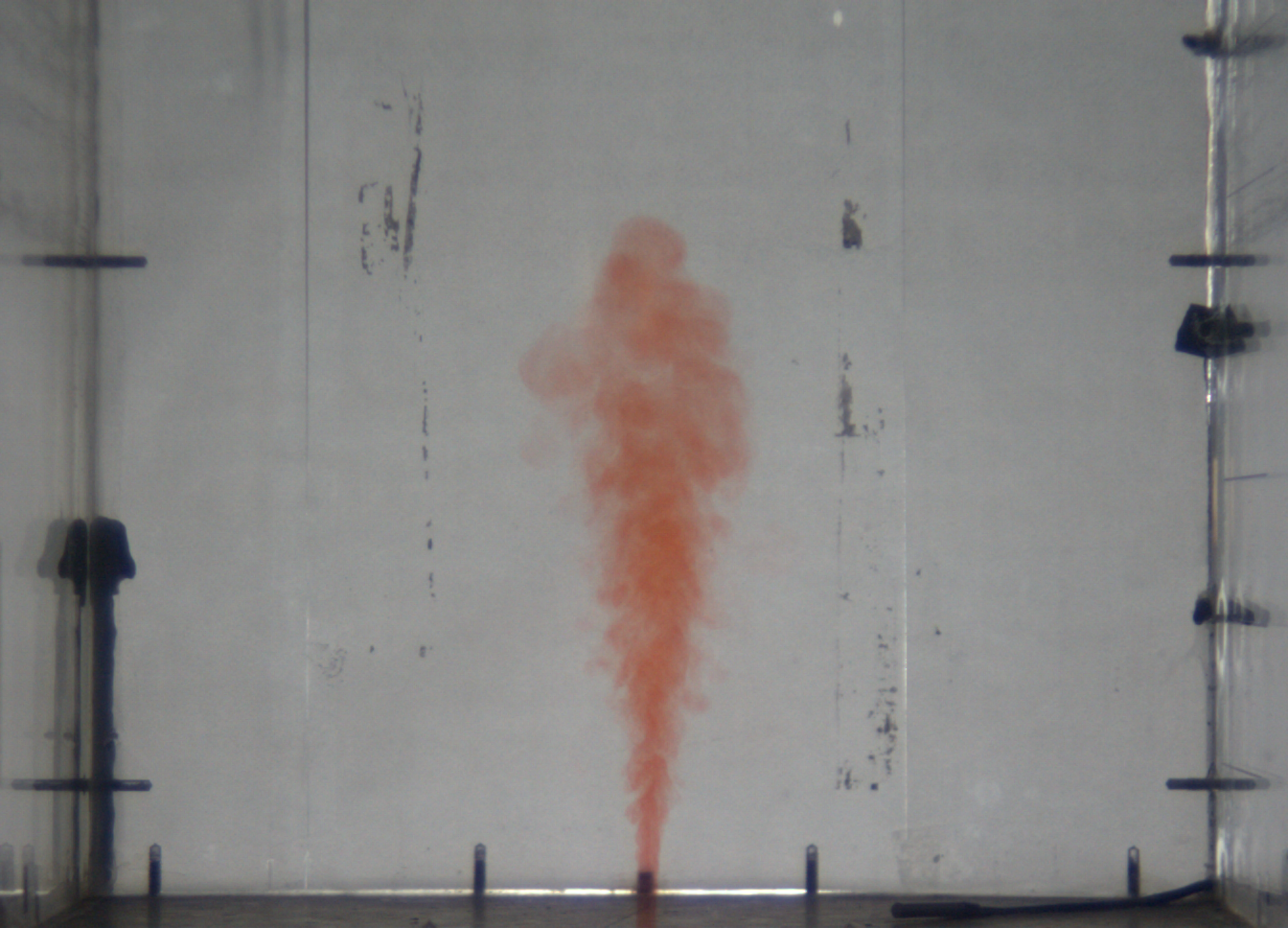}
		%\caption{View of $\phi$ over the $(x,z)$-plane.}
		\caption{}
		\label{fig:tinta_5s}
	\end{subfigure}
	%\hfill
	\hspace{0.5pc}
	\begin{subfigure}[b]{0.305\textwidth}
		\centering
		\includegraphics[width=\linewidth]{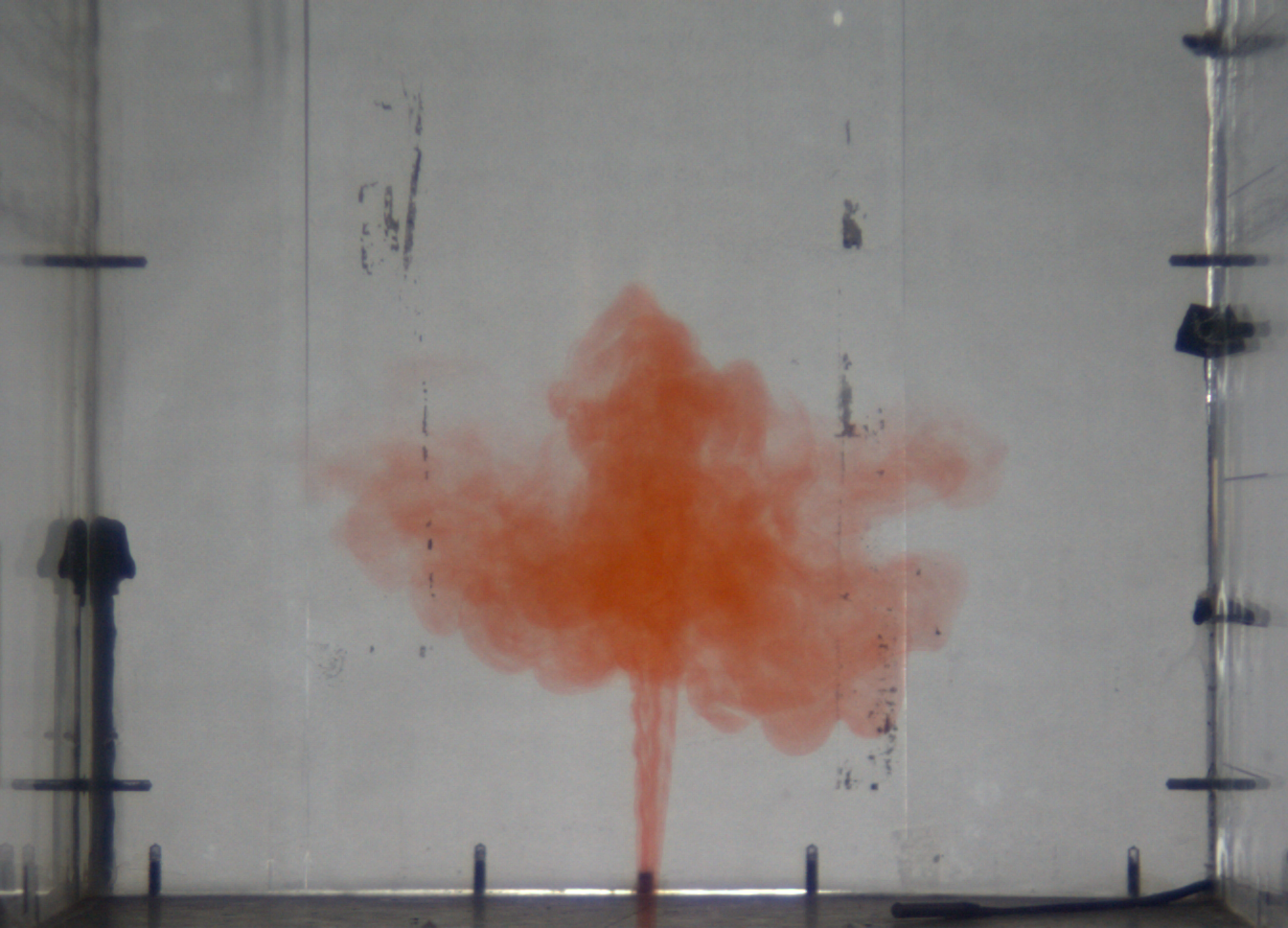}
		%\caption{View of $\phi$ over the $(x,y)$-plane.}
		\caption{}
		\label{fig:tinta_30s}
	\end{subfigure}
	\hspace{0.5pc}
	\begin{subfigure}[b]{0.305\textwidth}
		\centering
		\includegraphics[width=\linewidth]{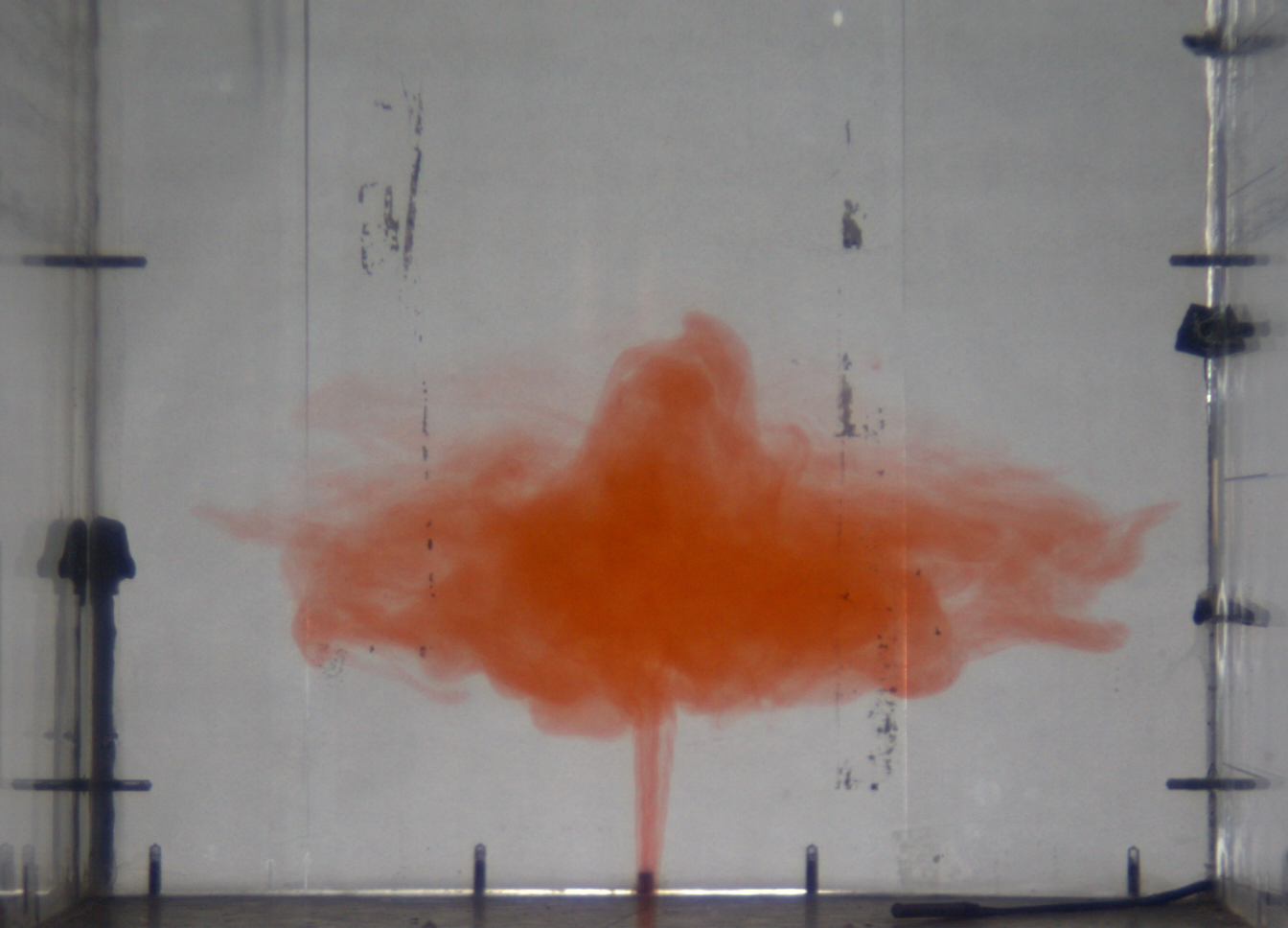}
		%\caption{View of $\phi$ over the $(x,y)$-plane.}
		\caption{}
		\label{fig:tinta_65s}
	\end{subfigure}
	\\
	\begin{subfigure}[b]{0.305\textwidth}
		\centering
		\includegraphics[width=\linewidth]{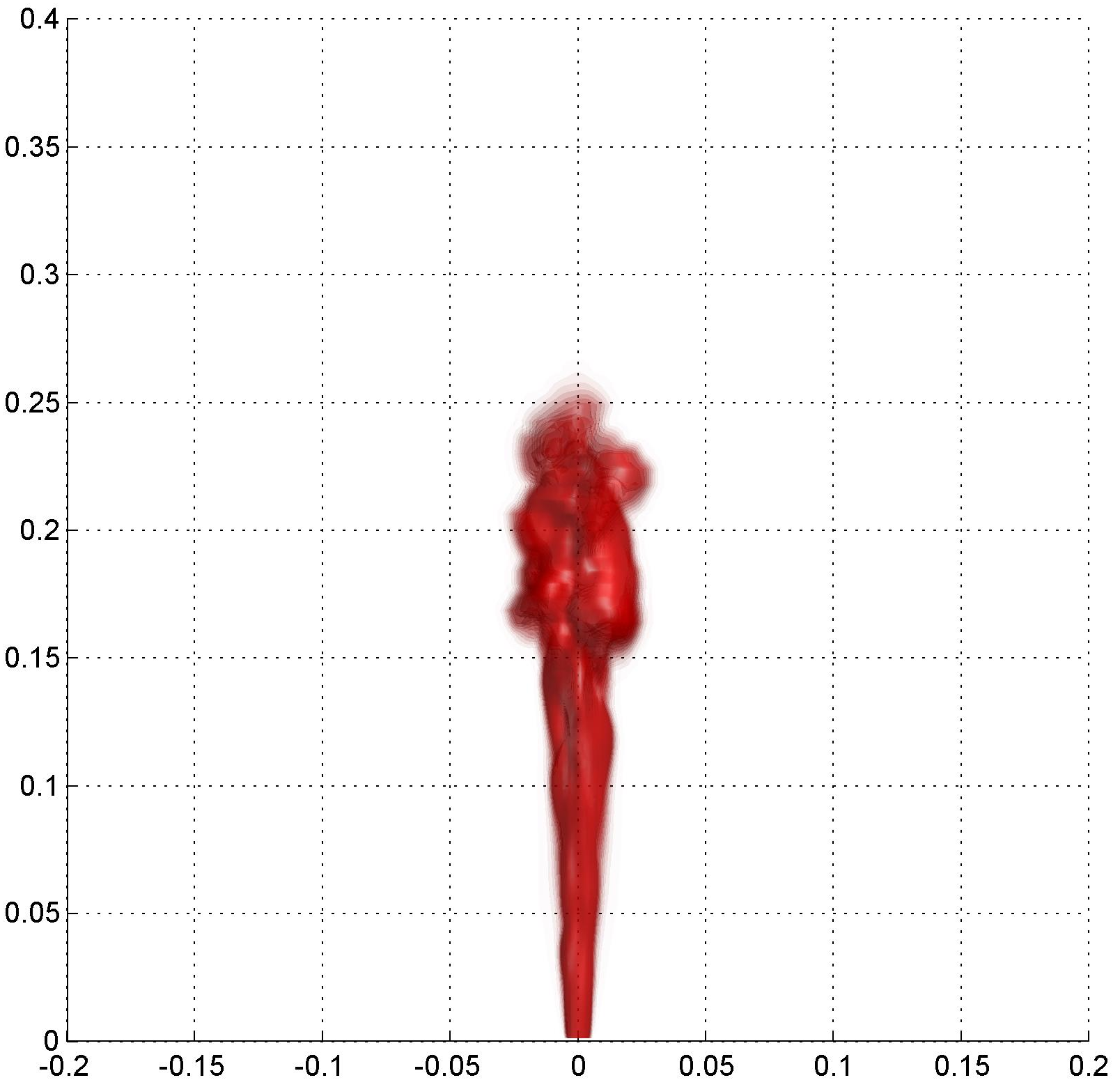}
		%\caption{View of $\phi$ over the $(x,z)$-plane.}
		\caption{}
		\label{fig:tracer_5s}
	\end{subfigure}
	%\hfill
	\hspace{0.5pc}
	\begin{subfigure}[b]{0.305\textwidth}
		\centering
		\includegraphics[width=\linewidth]{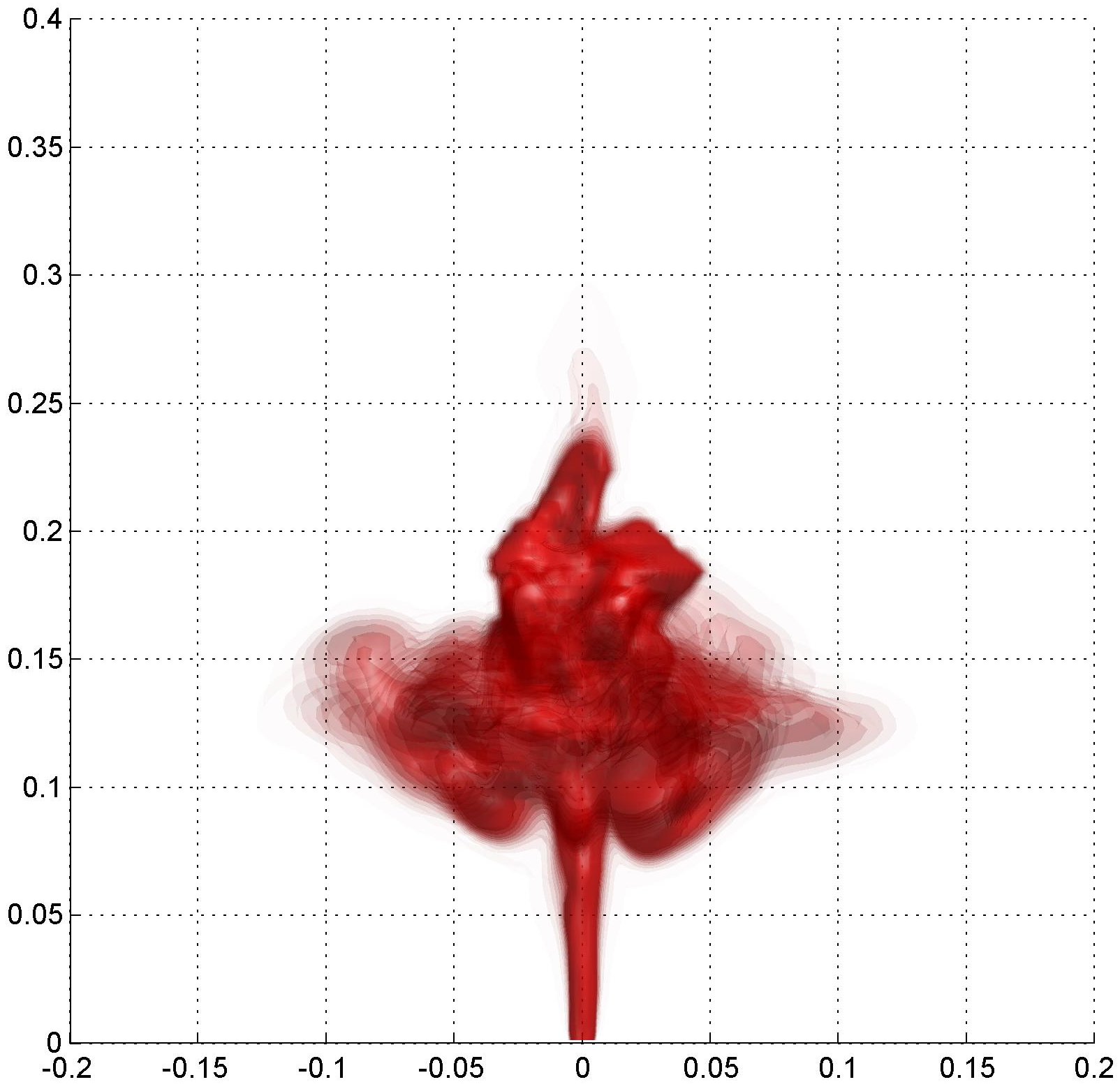}
		%\caption{View of $\phi$ over the $(x,y)$-plane.}
		\caption{}
		\label{fig:tracer_30s}
	\end{subfigure}
	\hspace{0.5pc}
	\begin{subfigure}[b]{0.305\textwidth}
		\centering
		\includegraphics[width=\linewidth]{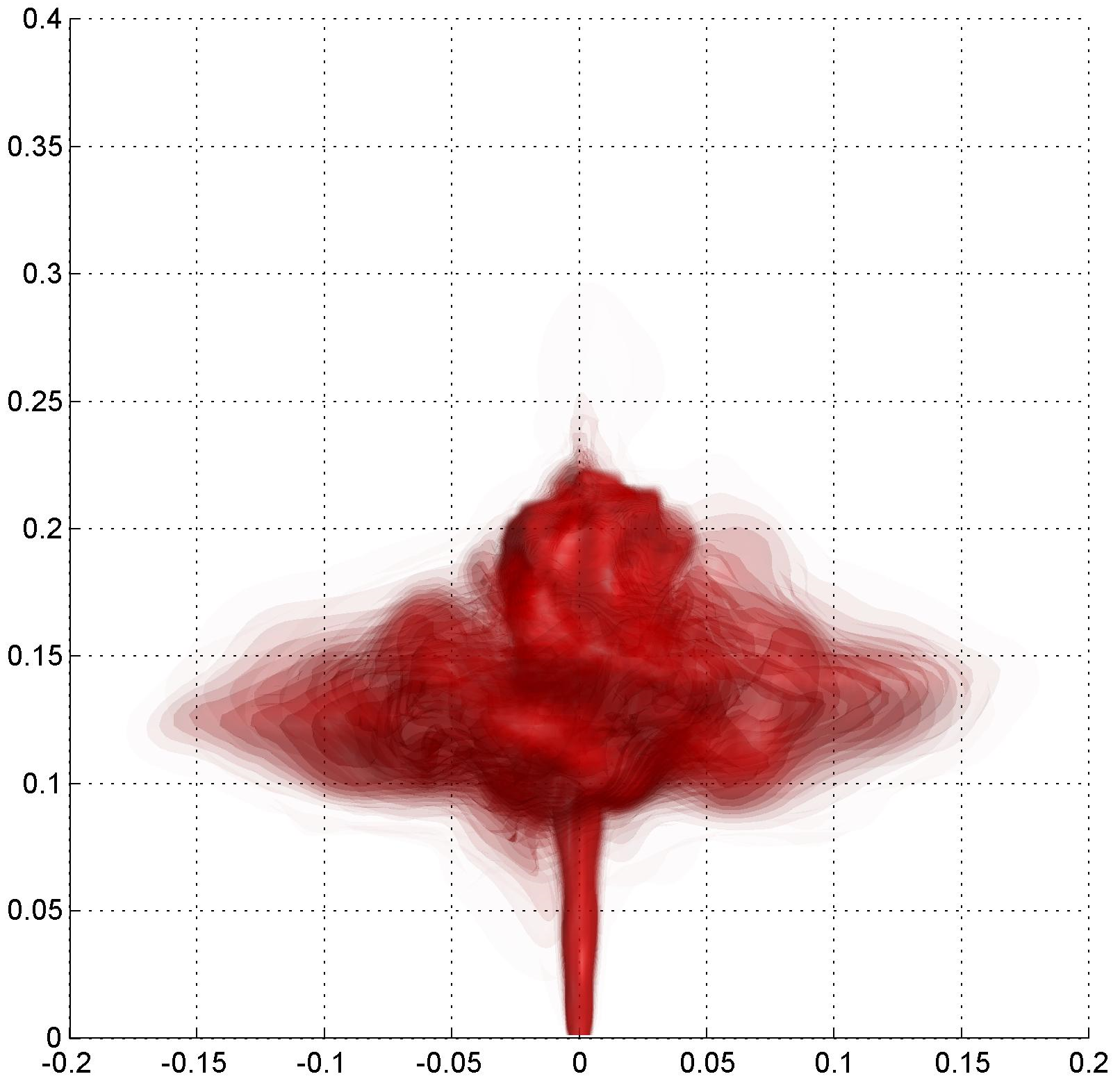}
		%\caption{View of $\phi$ over the $(x,y)$-plane.}
		\caption{}
		\label{fig:tracer_65s}
	\end{subfigure}
	
	\caption{ Experimental (first row) and numerical (second row)
          results obtained using the red tracer technique for the grid
          and $(\xi=0, u^\prime/U=20\%)$ configuration. Panels (a) and
          (d) correspond to $t = 5$~s, (b) and (e) to 30~s, and (c)
          and (f) to 65~s. The lower boundary of the first-row images
          represents the full width of the container (40~cm), which
          serves as the length-scale reference.  }
	\label{fig:tinta_experimental_3figs}
\end{figure}

%Fig.\ref{fig:tinta_sims_3figs}
Figs.~\ref{fig:tinta_experimental_3figs}(d)--(f) shows the numerical
results of $\phi$ for the $(\xi=0, u^\prime/U=20\%)$ configuration. As
can be easily observed by comparison with
Figs.~\ref{fig:tinta_experimental_3figs}(a)--(c), this configuration
corresponds to the experimental grid configuration. The same holds if
we compare the results of the free experimental configuration with
those of the numerical $(\xi=0, u^\prime/U=1\%)$ configuration.

\begin{figure}[htb!]
	\centering
	\includegraphics[width=0.55\linewidth,keepaspectratio]{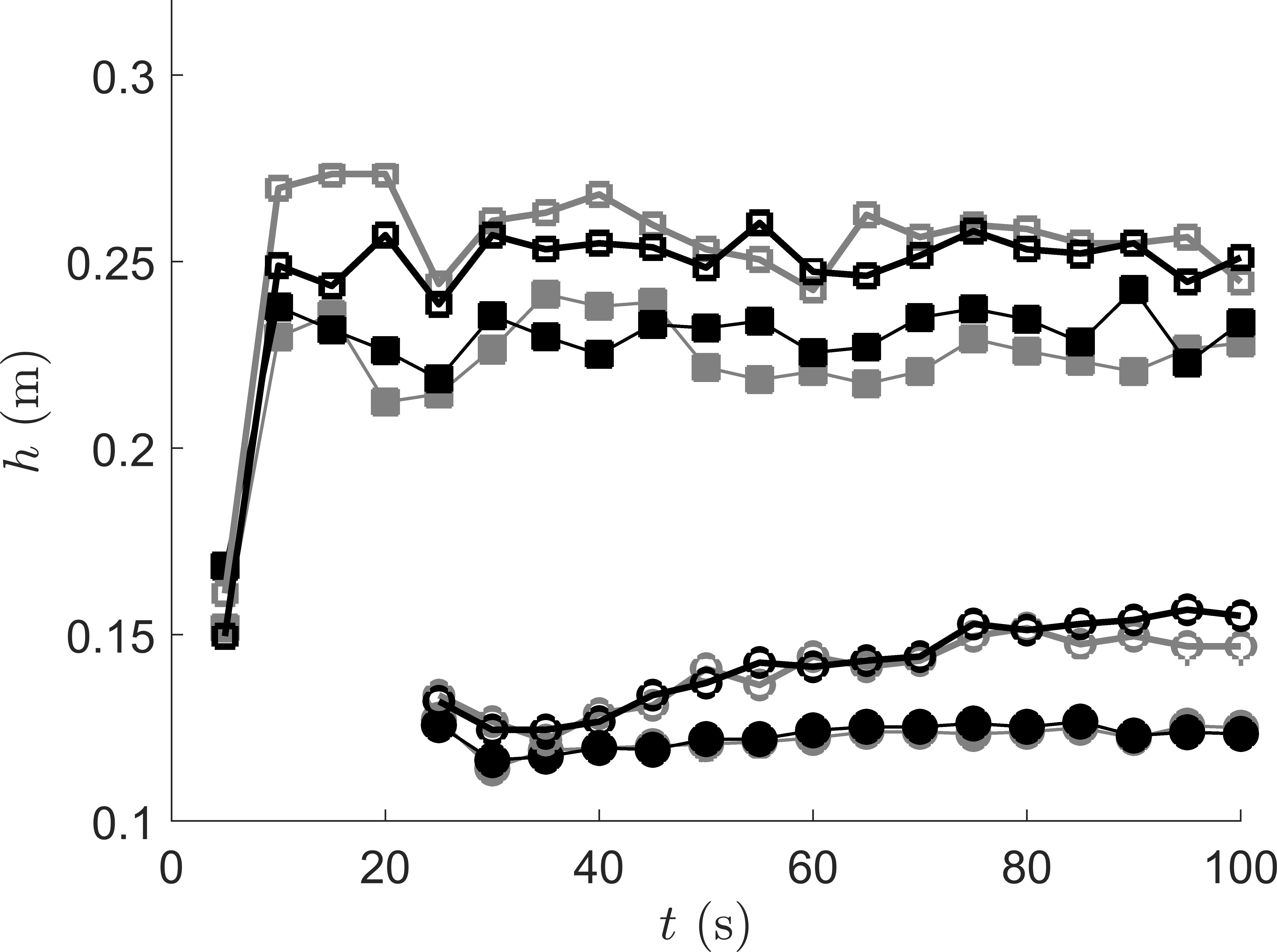}
	\caption{The experimental and numerical results for the
          maximum and spreading heights of the fountain are shown. The
          black (grey) markers correspond to the experimental
          (numerical) results. The square (circular) markers
          correspond to $h_m$ ($h_{sp}$), and the hollow (filled)
          markers correspond to the free (grid) configuration with
          $\xi=0$ and $u^\prime/U=1\%$ ($20\%$).}
	\label{fig:exp_and_sims_heights}
\end{figure}

One of the most significant analyses is the comparison of the time
evolution of the characteristic heights $h_m$ and $h_{sp}$ between
experiments and numerical simulations (further details about the
height measurements are given in
Sec.~\ref{subsec:height_measurement}). This analysis is presented in
Fig.\ref{fig:exp_and_sims_heights}. It confirms that the evolution of
the fountain for the free configuration follows the same pattern as
the numerical simulation for $u^\prime/U=1\%$, while the grid
configuration follows the numerical simulation for
$u^\prime/U=20\%$. The comparison of the characteristic heights
provides an essential basis for understanding the physical processes
involved in the fountain's evolution and helps to validate the
numerical simulations.

\begin{figure}[htb!]
	\centering
	\begin{subfigure}[b]{0.35\textwidth}
		\centering
		\includegraphics[height=\linewidth]{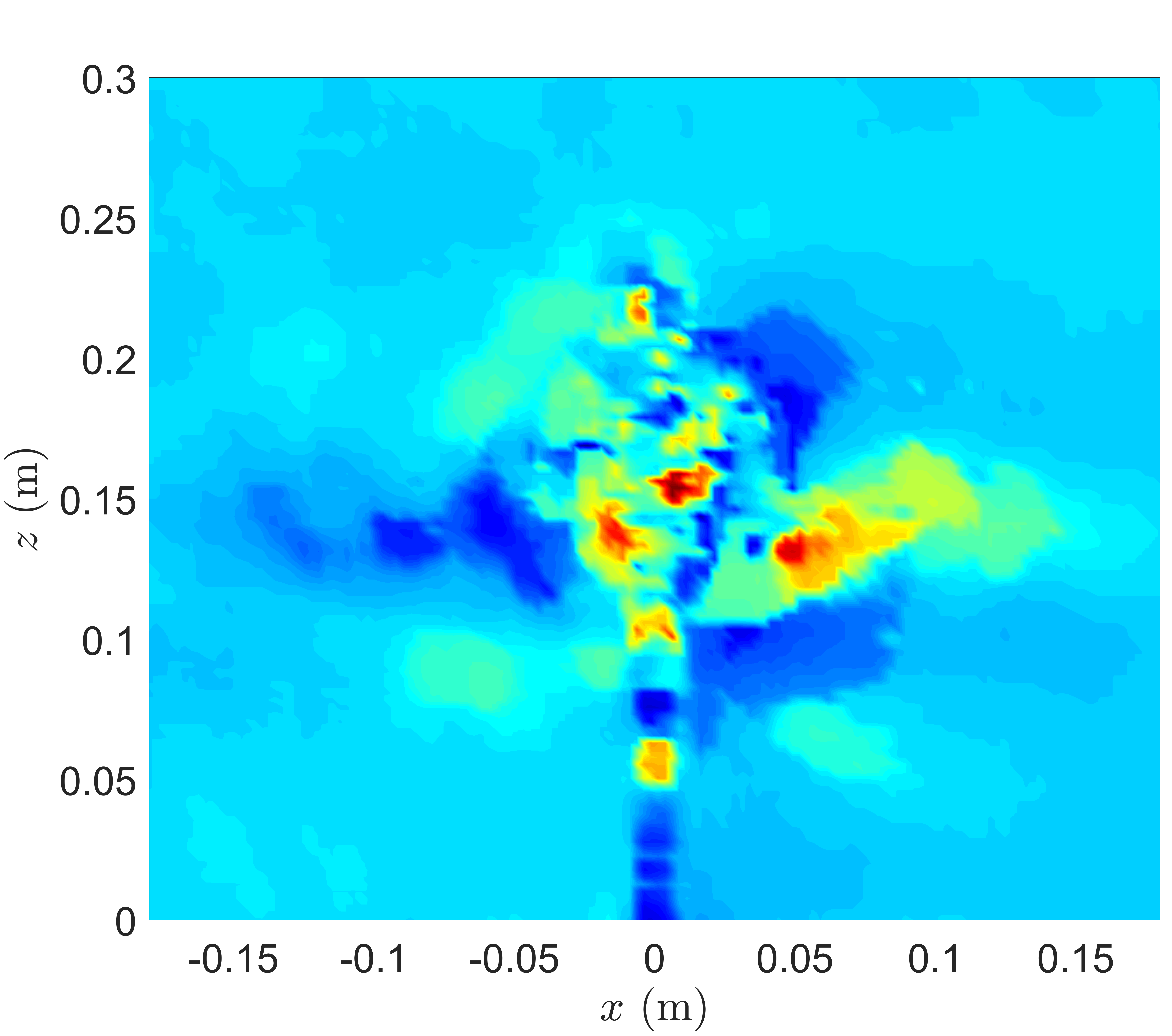}
		%\caption{View of $\phi$ over the $(x,z)$-plane.}
		\caption{}
		%\label{fig:constantforcesRA}
	\end{subfigure}
	%\hfill
	\hspace{1pc}
	\begin{subfigure}[b]{0.35\textwidth}
		\centering
		\includegraphics[height=\linewidth]{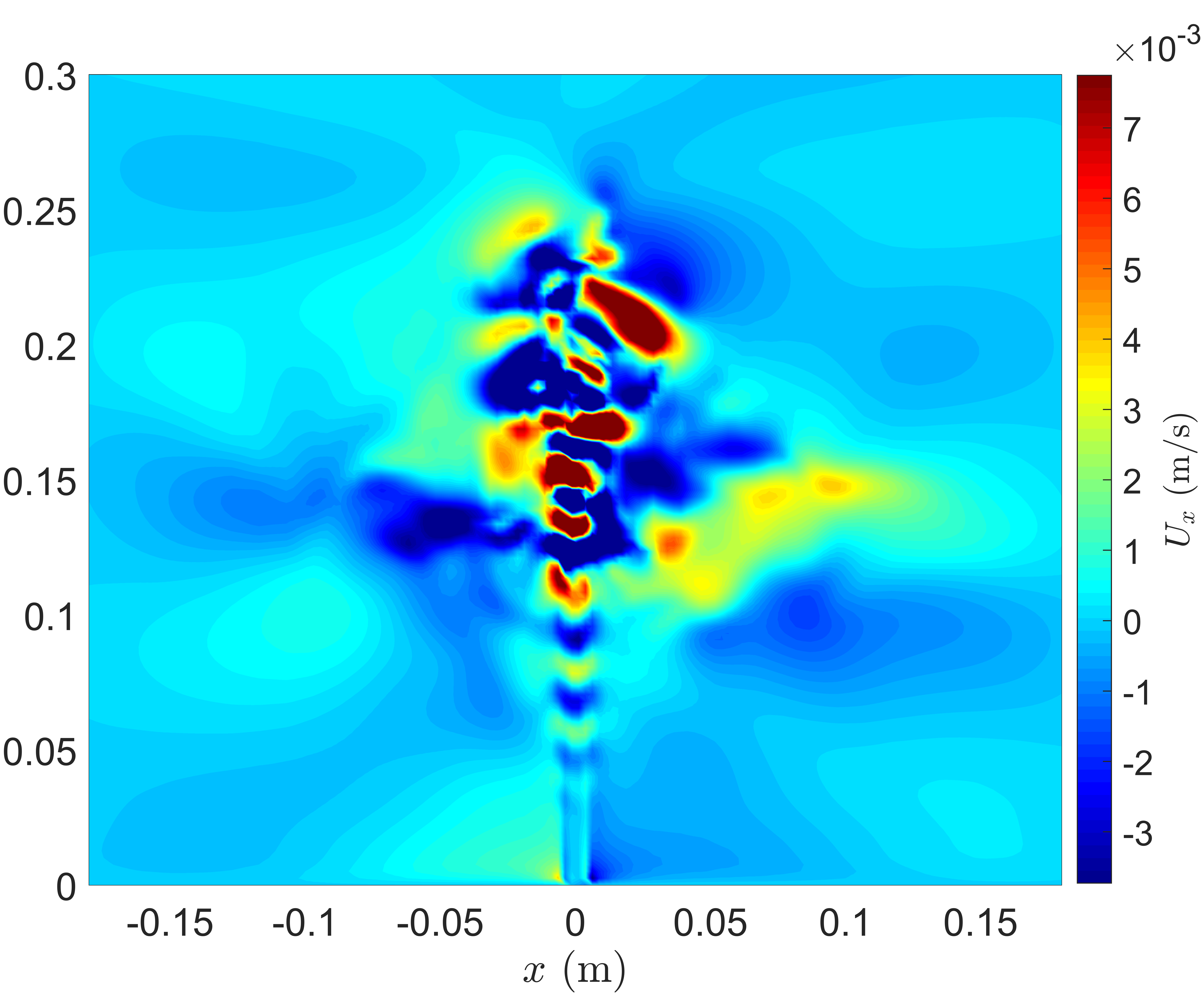}
		%\caption{View of $\phi$ over the $(x,y)$-plane.}
		\caption{}
		%\label{fig:sineforcesRA2}
	\end{subfigure}
\\
\centering
\begin{subfigure}[b]{0.35\textwidth}
	\centering
	\includegraphics[height=\linewidth]{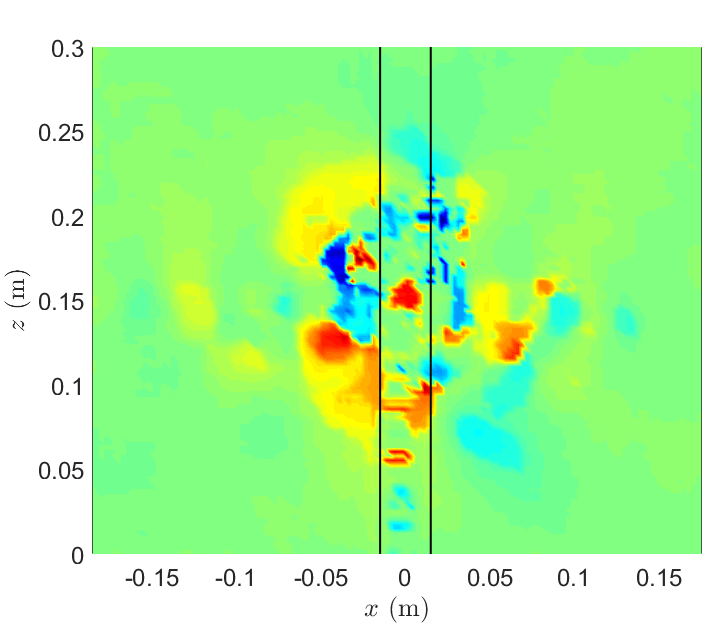}
	%\caption{View of $\phi$ over the $(x,z)$-plane.}
	\caption{}
	%\label{fig:constantforcesRA}
\end{subfigure}
%\hfill
\hspace{1pc}
\begin{subfigure}[b]{0.35\textwidth}
	\centering
	\includegraphics[height=\linewidth]{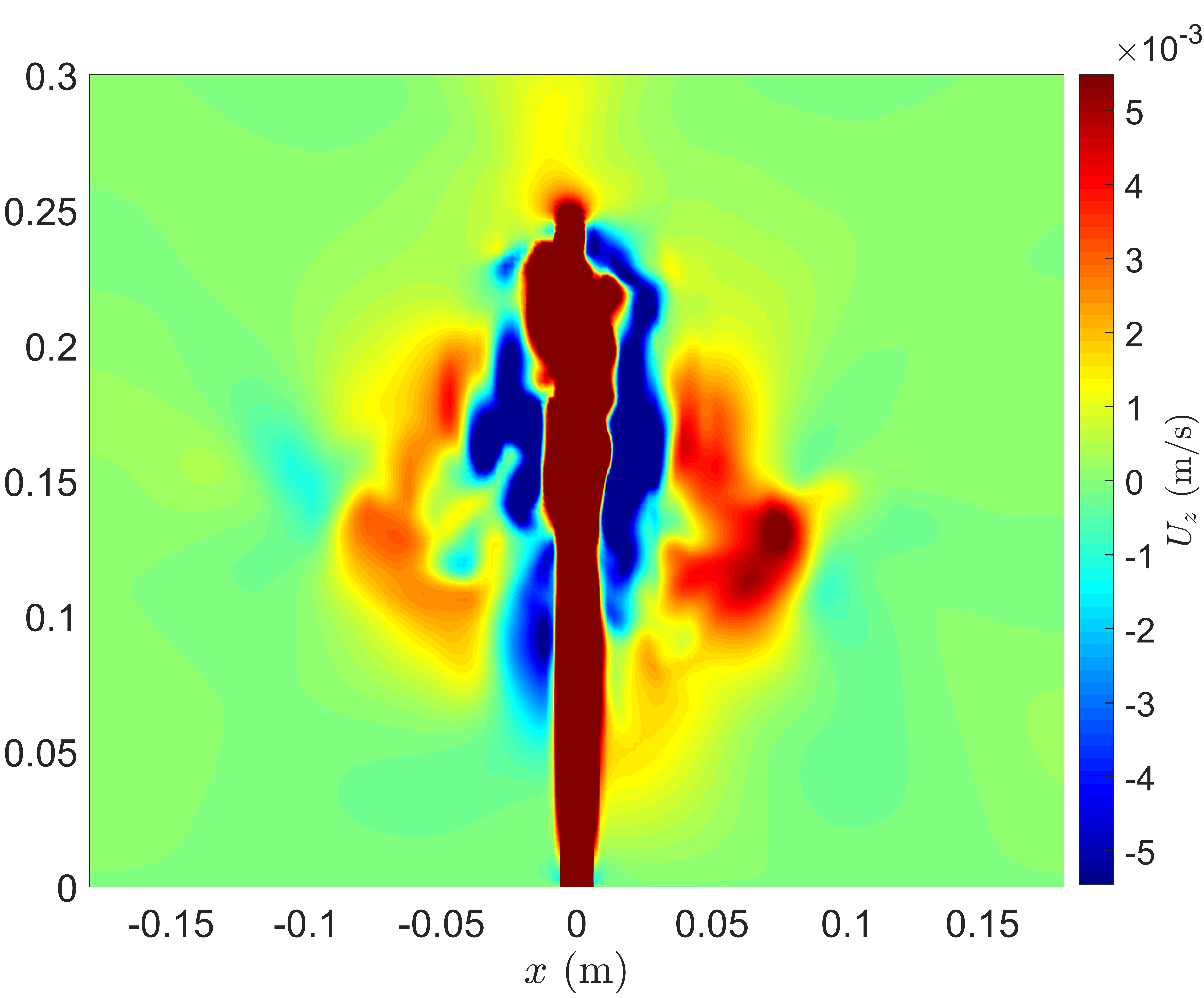}
	%\caption{View of $\phi$ over the $(x,y)$-plane.}
	\caption{}
	%\label{fig:sineforcesRA2}
\end{subfigure}
	\caption{Horizontal (top row) and vertical (bottom row)
          velocity components at 60 s obtained using the DPIV
          technique for the free experimental and numerical
          ($u^\prime/U=1\%$) configurations. The first column
          corresponds to the experimental data, and the second column
          corresponds to the numerical simulations. The color scales
          are consistent within each row. The central jet is not
          visible in (c) due to the limitations of the CMOS digital
          camera used. Therefore, comparison between (c) and (d) must
          be made outside the fountain axis region, delimited by the
          two vertical black lines in (c).}
	\label{fig:campos_vel}
\end{figure}

Finally, we compared the two-dimensional velocity fields measured
experimentally at the plane $y=0$ using the DPIV technique with the
results obtained from the numerical simulations. The corresponding
results for the free (grid) experimental configuration were found to
be in excellent agreement with the simulations for $\xi=0$ and
$u^\prime/U=1\%$ ($20\%$). For instance, Fig.~\ref{fig:campos_vel}
shows the velocity components obtained experimentally for the free
experimental configuration at $t=60$~s, and we compare the results
with the simulations performed for $\xi=0$ and $u^\prime/U=1\%$ at the
same time.

Based on the present analysis, we conclude that the simulations are in
good agreement with the experimental results, and therefore, we can
proceed with parameter sweeps across the proposed range.

%%%%%%%%%%%%%%%%%%%%%%%%
%\subsection{Numerical simulations} \label{sec:sims}
%%%%%%%%%%%%%%%%%%%%%%%%

\section{Results} \label{sec:results}

%, $N=\sqrt{-\frac{g}{\rho_{00}}\frac{d\rho_0}{dz}}$, where $\rho_0(z)$ is the initial density field of the stratified ambient fluid, $\rho_{00}=\rho_{0}(z=0)$ and $g=9.81$~$\mathrm{m}/\mathrm{s}^2$ is the gravitational acceleration, resulted $N = 0.2561$~s$^{-1}$.
%Provided that the thermal stratification was the explained in Sec.~\ref{sec:experimentos}, the Brunt-Vaisala frequency was fixed at $N = 0.2561$~s$^{-1}$ throughout all the calculations performed. The mean velocity field at the entrance is $\left< \mathbf{U_{in}} \right>=(0, 0, U)$, where $ \left\langle \cdot \right\rangle$ stands for the time average and the flow rate was set as $\dot{q}_{in}=5.5$~$\mathrm{cm}^3/s$ for all the cases in the present work.  In the laboratory experiments, we used $T_{in}=15^{\circ}$C, meanwhile in the numerical simulations, we considered $T_{in}=4$, $5$, $7.5$, $10$, $12.5$ and $15^{\circ}$C. 

After validating the numerical results, simulations were conducted to
explore the effect of varying values of the temperature of the
fountain at the inlet and turbulence intensity. Specifically,
simulations were performed for all combinations of $T_{in}$ = 4, 5,
7.5, 10, 12.5, and 15$^{\circ}$C and $u^{\prime}/U = 0$, $1$, $2$,
$4$, $10$, and $20\%$, respectively. To quantify our findings, we
introduce the dimensionless parameter $\xi$. This parameter is defined
as $\xi=-\mathrm{Fr}^{-2}$, where $\mathrm{Fr}$ represents the Froude
number at the inlet, which can be expressed as $\mathrm{Fr}=U/\sqrt{g
  D \frac{\rho_{in}-\rho_{00}}{\rho_{00}}}$,
%\begin{equation}
%	\centering
%	\displaystyle
%	\mathrm{Fr}=\frac{U}{\sqrt{g D \frac{\rho_{in}-\rho_{00}}{\rho_{00}}}} ,
%	\label{eq:froude_numer}
%\end{equation}
where $\rho_{in}$ is the density of the fountain at the inlet, $\rho_0(z)$ represents the initial density field of the stratified ambient fluid, and $\rho_{00}=\rho_{0}(z=0)$ denotes the density of water at the given temperature. The aforementioned densities were computed based on the fluid temperature and tabulated properties of water.
Note that $\xi$ grows monotonically with $T_{in}$ since $\rho_{in}\leq \rho_{00}$, as shown in Tab.~\ref{tab:tjet_to_xi}. For this reason, we refer to $\xi$ as the \textit{lightness} of the fountain. 

%Note that although the lightness is related to the Richardson number as $\xi=-\mathrm{Ri}$, for the sake of clarity, we refer to the first dimensionless number instead of the later to quantify the incoming jet's inertia with respect to the gravitational force of the surrounding environment. The findings of this investigation provide empirical support for the proposition that $h_m$ increases as the jet exhibits greater lightness than the encompassing medium.
%Since $U$, $D$, $\rho_{00}$ and $\rho_{in}\leq \rho_{00}$ in the present work, $\xi$ grows monotonically with $T_{in}$, as shown in Tab.~\ref{tab:tjet_to_xi}.
%Note that although the lightness is related to the Richardson number as $\xi=-\mathrm{Ri}$, for the sake of clarity, we refer to the first dimensionless number instead of the later to quantify the incoming jet's inertia with respect to the gravitational force of the surrounding environment. The findings of this investigation provide empirical support for the proposition that $h_m$ increases as the jet exhibits greater lightness than the encompassing medium.

\begin{table}[htb!]
	\centering
	\caption{Fountain lightness, $\xi$,  as a function of the inlet temperature, $T_{in}$.}
	\begin{tabular}{l|rrrrrrr}
		$T_{in}$ ($^{\circ}C$)   & 4.0  & 5.0  & 6.0  & 7.5  & 10.0 & 12.5 & 15.0 \\ \hline
		$\xi$ ($\times 10^{-3}$) & -6.1 & -6.0 & -5.9 & -5.5 & -4.3 & -2.5 & 0   
	\end{tabular}
	\label{tab:tjet_to_xi}
\end{table}

\begin{figure}[htb!]
	\centering
	\begin{subfigure}[b]{0.305\textwidth}
		\centering
		\includegraphics[width=\linewidth]{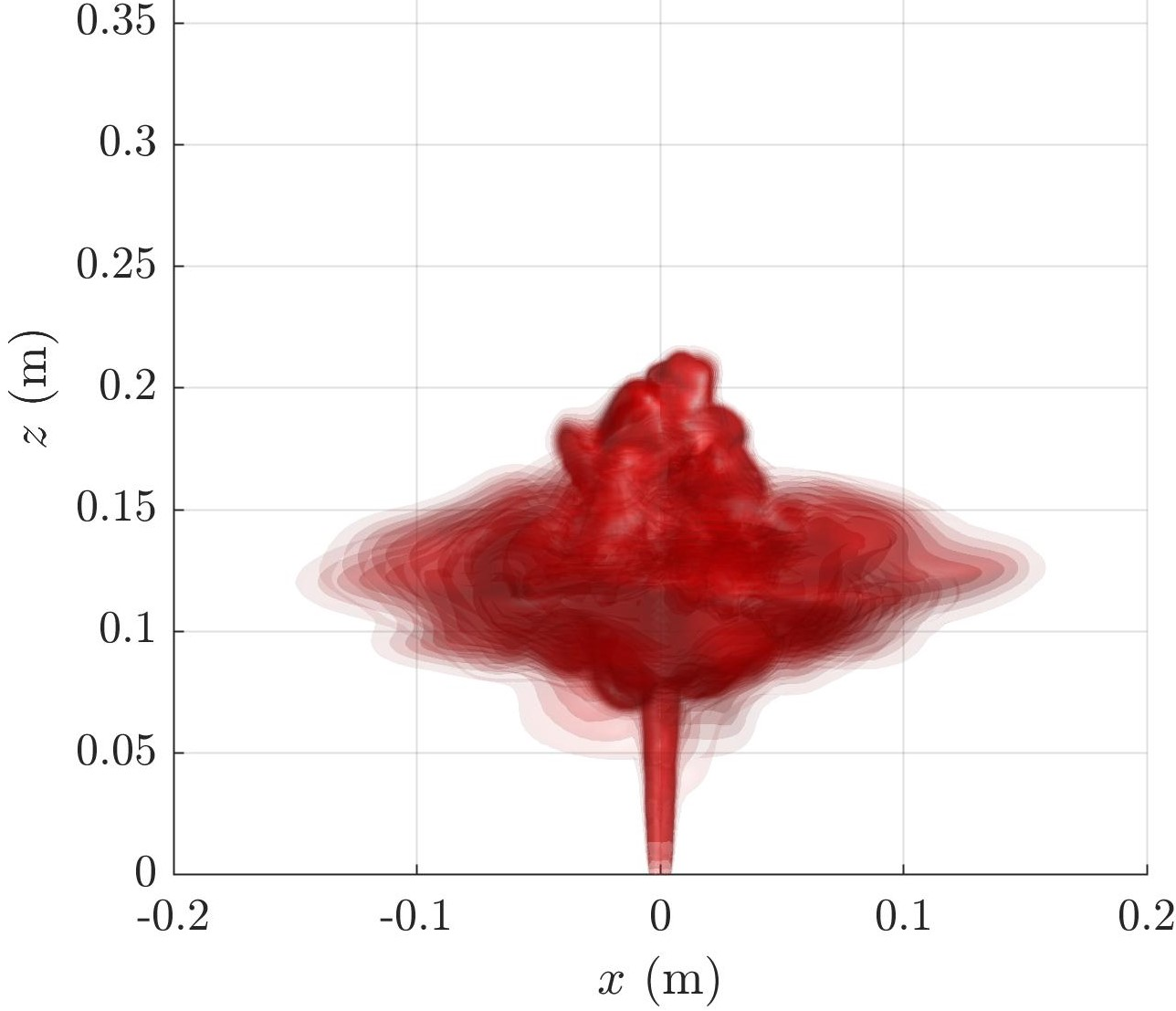}
		%\caption{View of $\phi$ over the $(x,z)$-plane.}
		\caption{}
		\label{fig:tracer_nocollapse}
	\end{subfigure}
	%\hfill
	\hspace{0.5pc}
	\begin{subfigure}[b]{0.305\textwidth}
		\centering
		\includegraphics[width=\linewidth]{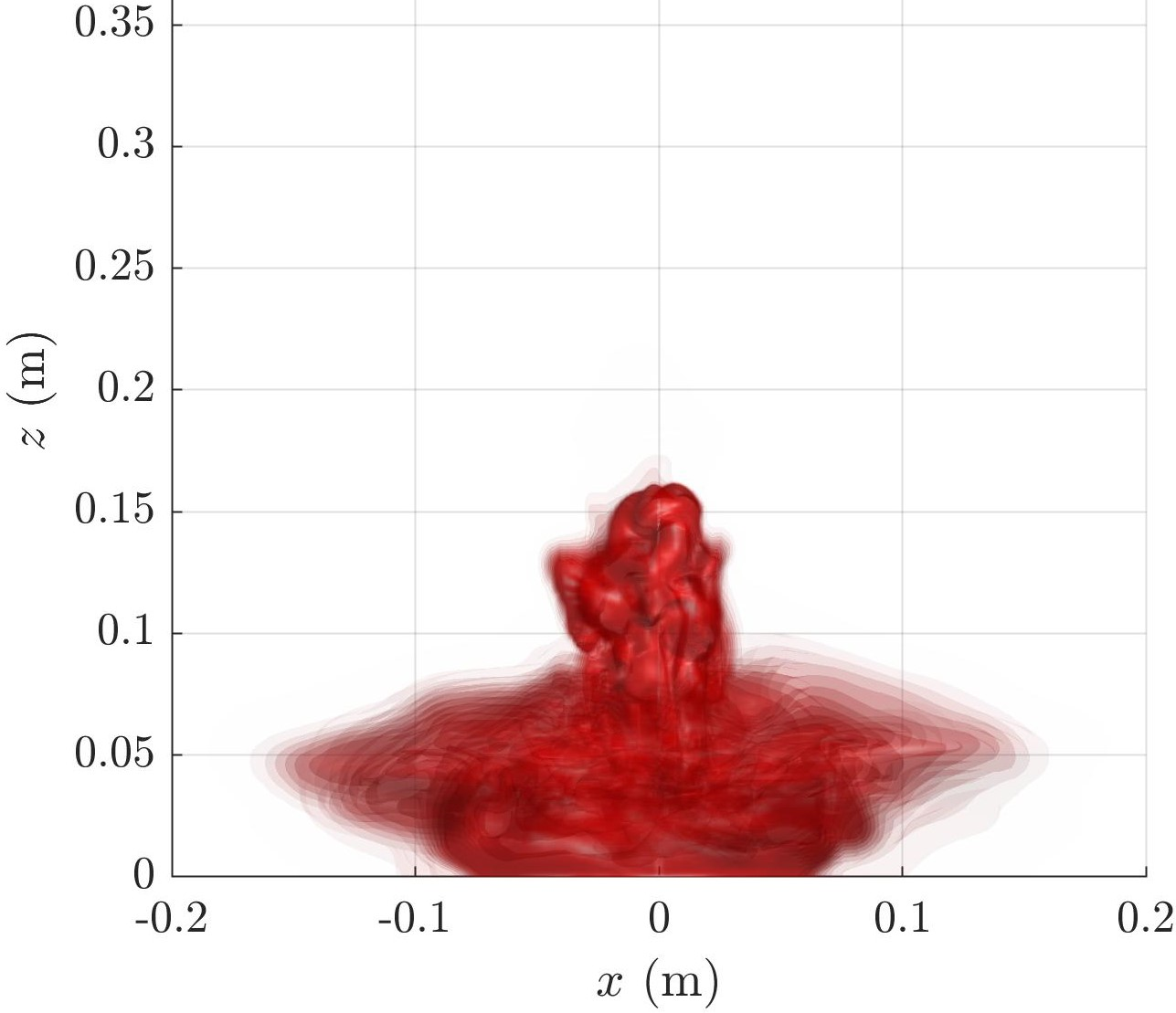}
		%\caption{View of $\phi$ over the $(x,y)$-plane.}
		\caption{}
		\label{fig:tracer_semicollapes}
	\end{subfigure}
	\hspace{0.5pc}
	\begin{subfigure}[b]{0.305\textwidth}
		\centering
		\includegraphics[width=\linewidth]{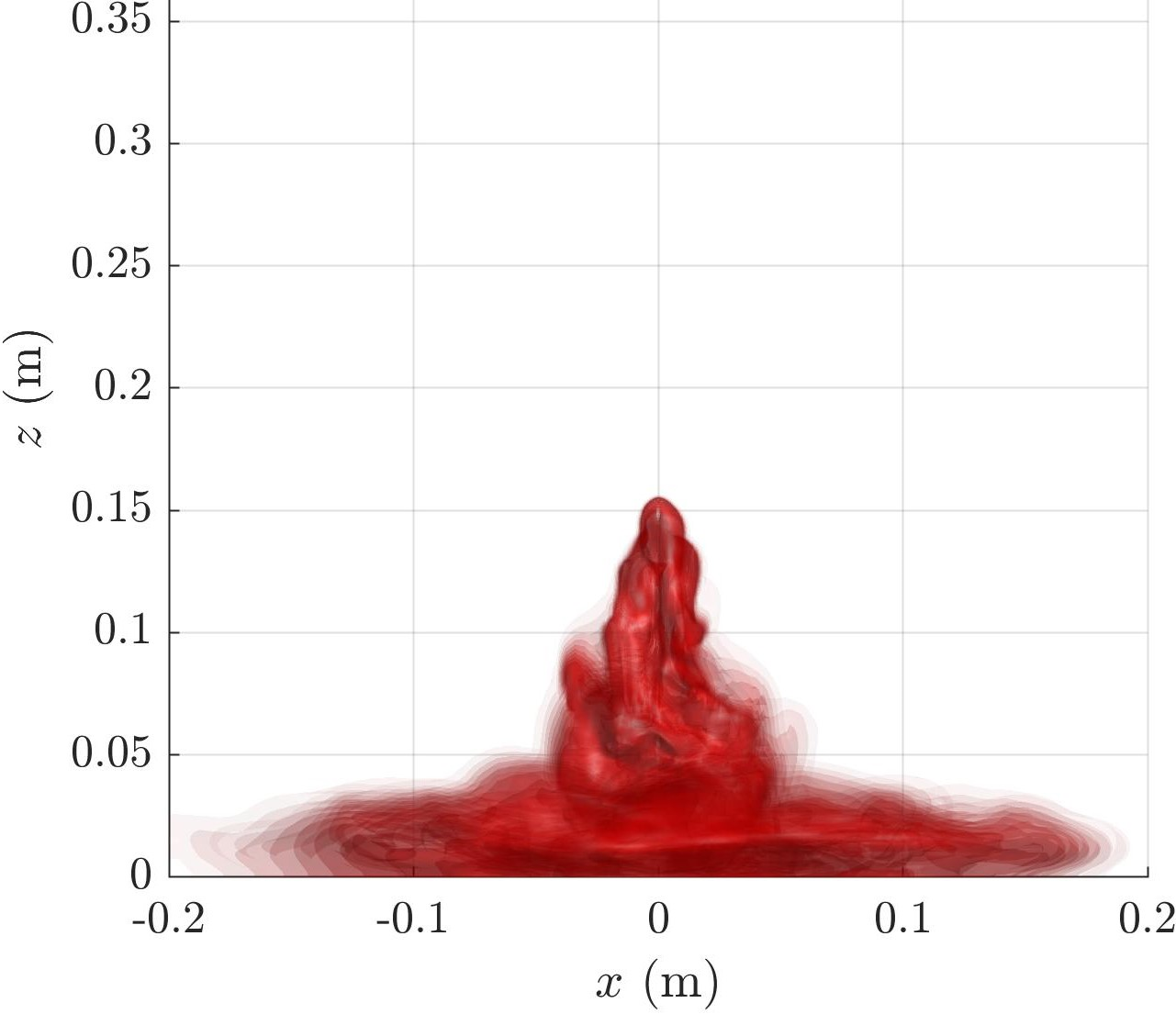}
		%\caption{View of $\phi$ over the $(x,y)$-plane.}
		\caption{}
		\label{fig:tracer_collapse}
	\end{subfigure}
	\caption{ Passive tracer field for the three different regimes
          of the fountain dynamics observed: (a) no-collapse
          ($\xi=0$), (b) semi-collapse ($\xi=-4.3\times10^{-3}$), and
          (c) collapse ($\xi=-6.1\times10^{-3}$). Parameter value:
          $u^\prime/U=20\%$.}
	\label{fig:tracer_different_Froudes}
\end{figure}

Our study aimed to analyze the flow characteristics resulting from
different combinations of two control parameters: the lightness and
turbulence intensity of the fountain.  In
Fig.~\ref{fig:tracer_different_Froudes}, we present numerical results
for the passive tracer field with a turbulence intensity of $20\%$ and
three distinct lightness values, leading to the three different
regimes described earlier.

%%%%%%%%%%%%%%%%%%%%%%%

%%%%%%%%%%%%%%%%%%%%%%

In this section, we organize the discussion as follows: in
Sec.~\ref{subsec:height_measurement}, we provide details of the
procedure used to obtain the characteristic heights of the flow from
the numerical simulations. In Sec.~\ref{sec:alturas}, we present the
results and corresponding analysis. Finally, in
Sec.~\ref{sec:terreno}, we examine the impact of the
fountain-developed dynamics on the effectiveness of its technological
applications for contaminant removal, specifically, the Selective
Inverted Sink (SIS) device.

\subsection{Characteristic heights measurement}
\label{subsec:height_measurement}

\begin{figure}%[tb!]
    \centering
    \includegraphics[width=0.55\linewidth,keepaspectratio]{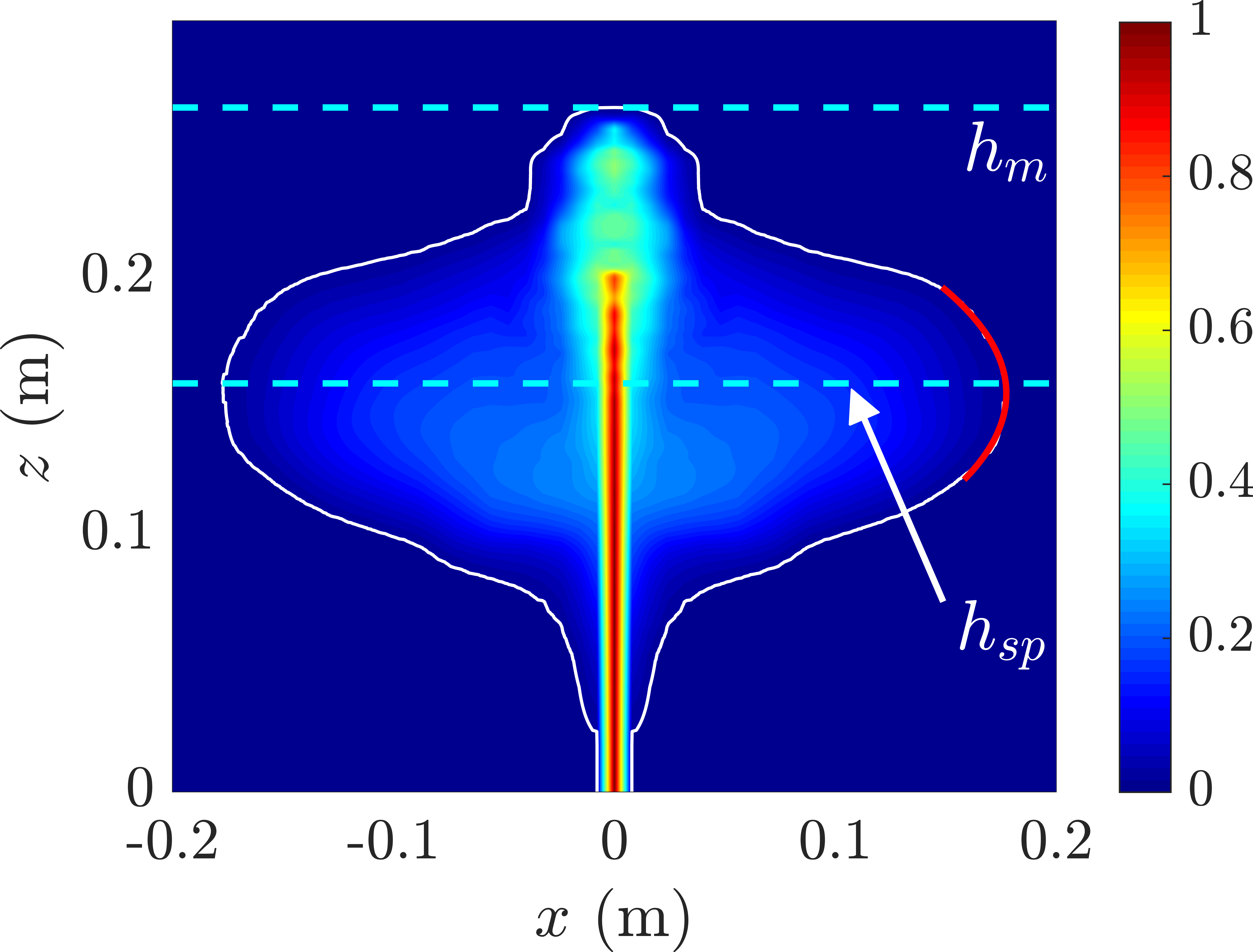}
    \caption{ Colour map of the azimuthal average of the tracer
      concentration ($\phi$) at $t=110$~s for the configuration
    %$(\xi=0,\frac{u^{\prime}}{U}=1\%)$
    %$(\xi=0,u^{\prime}/U=1\%)$
    $\xi=0$, $u^{\prime}/U=1\%$. The solid white line represents the
    contour of the averaged fountain $\mathscr{C}$. The top of such a
    contour defines $h_m$ (top cyan dashed horizontal line);
    meanwhile, the height of the extrema of the red line,
    corresponding to the parabola that fits the spreading cloud front,
    corresponds to $h_{sp}$ (bottom cyan dashed horizontal line).  }
    \label{fig:procesa_Tr}
\end{figure}

Given the complexity of the three-dimensional flow the measurement of
the characteristic heights $h_{m}$, $h_{sp}$ and $h_{c}$ requires the
proper processing
%requires the subtle manipulation
of the numerical results. In all cases, the calculation is based on
the passive scalar tracer concentration field $\phi$. Nevertheless, we
separate the strategy for the calculation of $h_{m}$ and $h_{sp}$
%(Sec.~\ref{ssseq:hm_y_hsp}) 
from the used for $h_c$.  We analysed the concentration field $\phi$
of the passive scalar tracer during the fully developed flow stage,
i.e., after the values of $h_m$ and $h_{sp}$ reached nearly constant
values.  %Specifically,
%this occurred at $t=100$~s.  Calculations on the characteristic
%heights were  averaged taken values  in the interval between 100 and $120$~s.
To define the contour of the developed fountain at each time
$\mathscr{C}(t)$, we initially established a threshold $\phi_{tol}$ in
an arbitrary manner.  After some experiments, we found
$\phi_{tol}=0.01$ sharply defines the fountain contour. The described
procedure applied to the tracer field from
Fig.~\ref{fig:3d_images_fountain} leads to the results shown in
Fig.~\ref{fig:procesa_Tr}. On one hand, we define $h_m(t)$ as the
height of  the top of $\mathscr{C}(t)$. On the other hand,
after approximating the front of the spreading cloud, at the maximum
radial position of $\mathscr{C}(t)$, with a horizontal parabola, we
obtain $h_{sp}(t)$ as the height of the extrema  of such a
parabola.  Finally, for a given configuration of lightness and
turbulence level of the fountain $(\xi, u^{\prime}/U)$, $h_m$ and
$h_{sp}$ are defined as the mean of $h_{m}(t)$ and $h_{sp}(t)$,
respectively, for $t=100$, $105$, $110$, $115$ and $120$~s.  The
dimensionless maximum and spreading heights, defined as
$h_m^{\ast}=h_m/D$ and $h_{sp}^{\ast}=h_{sp}/D$, are
shown in Figs.~\ref{fig:hmax} and \ref{fig:hsp}, respectively.

%-------------------------------------
%-------------------------------------

%\subsubsection{Measurement of $h_c$} \label{ssseq:hc}
To obtain $h_c$, we took into account that the grid is composed
ofhorizontal layers of cells. For each height, $z_c$, we computed
$\phi_{min}(z_c)$ as the minimum of $\phi$ for locations away from the
uprising fountain, i.e., beyond an arbitrary given cut-off radial
distance from the fountain axis. Such radial distance was chosen as
three times the radius of the inlet.  For a given tolerance value for
tracer concentration, $\phi_{tol}$, we define
\begin{equation}
\centering
%\displaystyle h_c(\phi_{tol})=\min \{z_c \mathrm{, \ s.t. \ } \phi_{min}(z_c)\geq\phi_{tol}\}.
\displaystyle h_c(\phi_{tol})=\min \{z_c \mathrm{, \ such \ that \ } \phi_{min}(z_c)\geq\phi_{tol}\}
%$\displaystyle h_c(\phi_{tol})=\min_{z_c \mathrm{ \ calculation \ cell}} \{z_c \mathrm{, \ such \ that \ } \phi_{min}(z_c)\geq\phi_{tol}\}$.
\label{eq:hc_phi_tol}
\end{equation}
and dimensionless critical height as
$h_c^{\ast}=h_c/D$.

\subsection{Results for the characteristic heights}
\label{sec:alturas}

\begin{figure}[htb!]
	\centering
	\begin{subfigure}[b]{0.49\textwidth}
 \includegraphics[width=\linewidth]{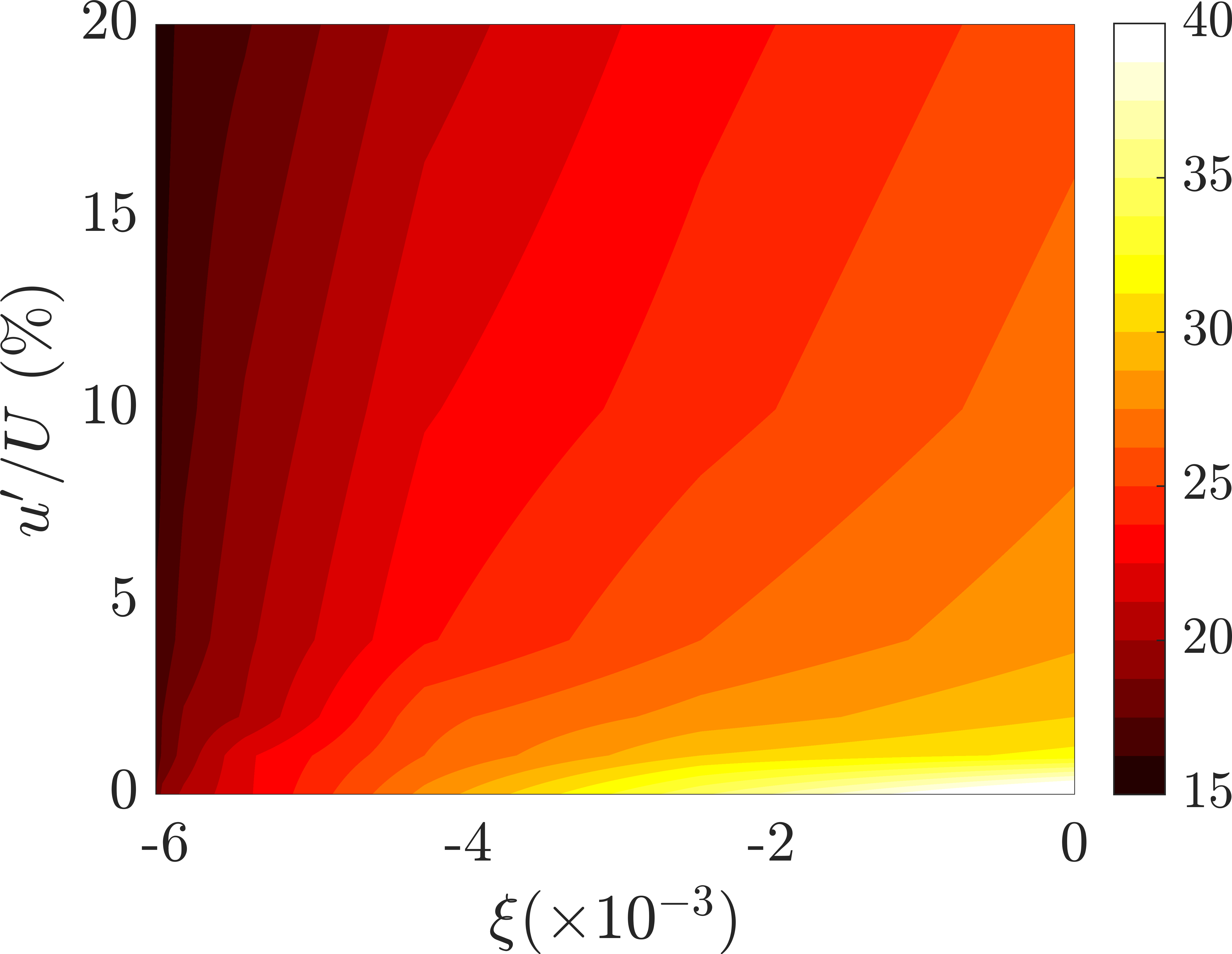}
		\caption{}
        %\caption{Heatmap of $h_m^{\ast}$.}
		\label{fig:hmax_calor}
	\end{subfigure}
	\hfill
	\begin{subfigure}[b]{0.49\textwidth}
 \includegraphics[width=\linewidth]{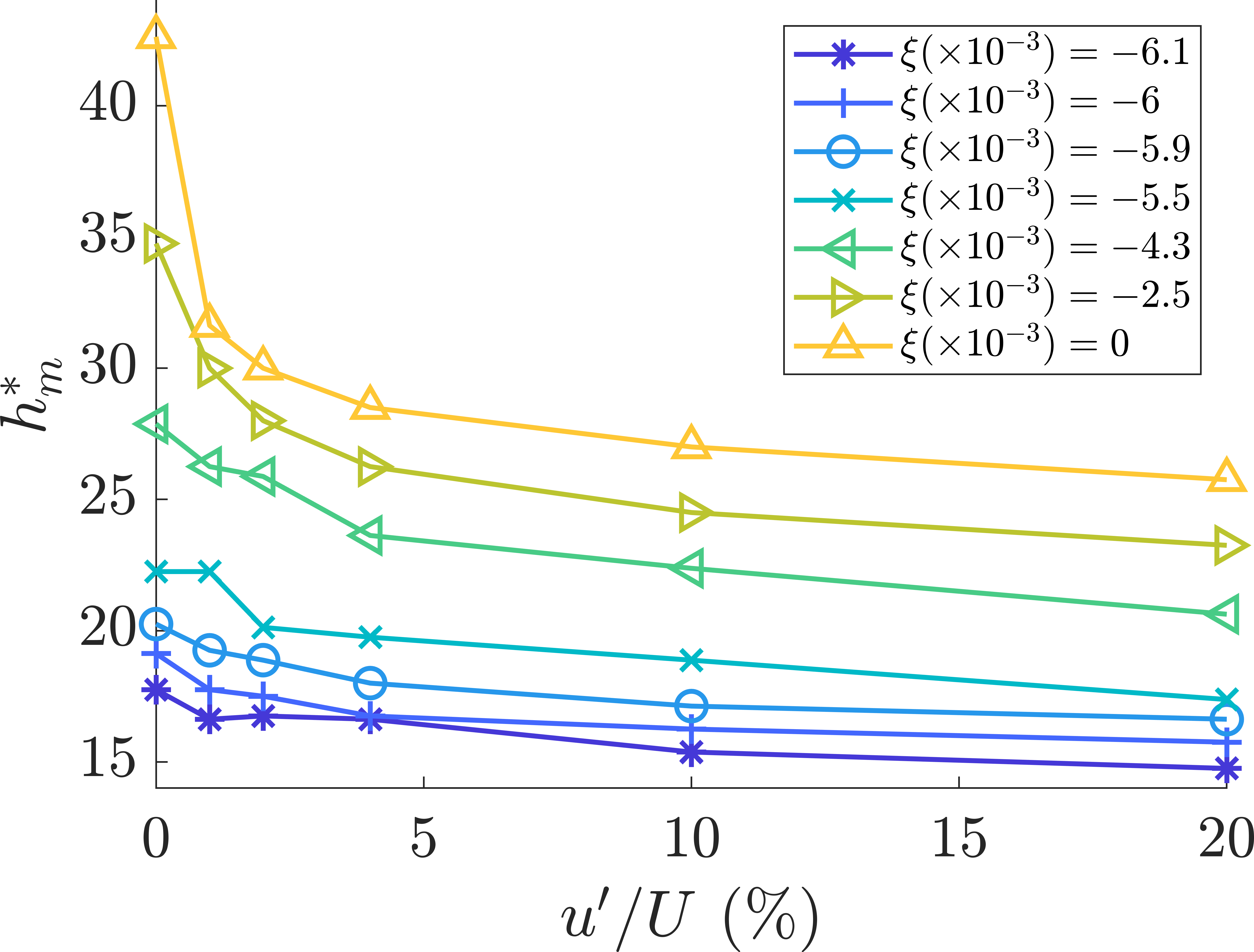}
    \caption{}
        %\caption{Level curves of $h_m^{\ast}$ at different values of $\xi$.}
		\label{fig:hmax_plots}
	\end{subfigure}
 %\caption{Two representations of the dimensionless maximum height reached by the fountain
 %%($h_m^{\ast}$)
 %as a function of its lightness ($\xi$) and turbulence level ($u^{\prime}/U$): (a) heatmap of $h_m^{\ast}$ in the phase space ($\xi$,$u^{\prime}/U$); (b) level curves of $h_m^{\ast}$ as a function of $u^{\prime}/U$ for different values of $\xi$.}
 \caption{Dimensionless maximum height, $h_{m}^{\ast}$, as a function
   of its lightness, $\xi$, and turbulence level, $u^{\prime}/U$: (a)
   heatmap; (b) level curves.}
 \label{fig:hmax}
\end{figure}

\begin{figure}[htb!]
	\centering
	\begin{subfigure}[b]{0.49\textwidth}
 \includegraphics[width=\linewidth]{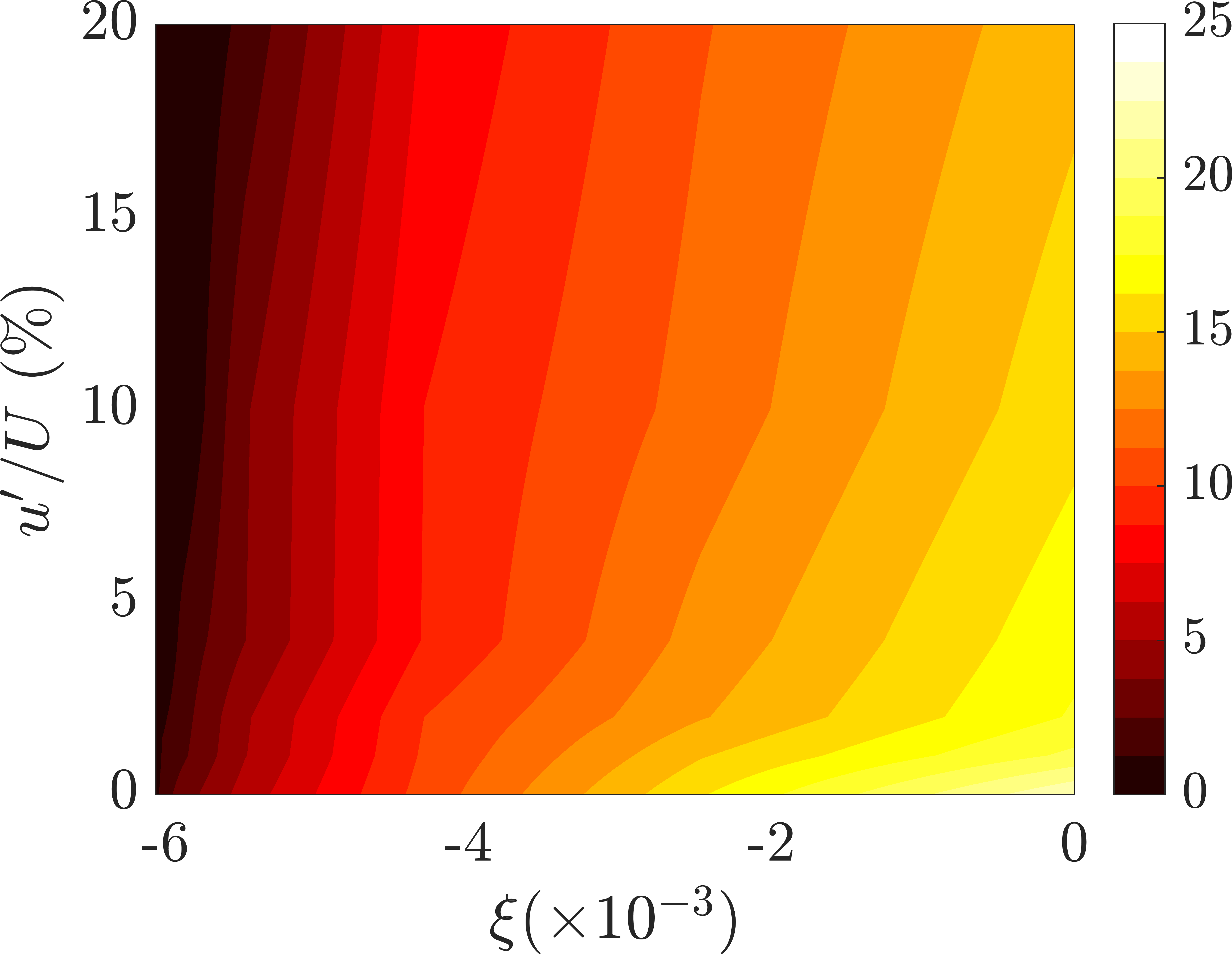}
		\caption{}
        %\caption{Heatmap of $h_{sp}^{\ast}$.}
		\label{fig:hsp_calor}
	\end{subfigure}
	\hfill
	\begin{subfigure}[b]{0.49\textwidth}
 \includegraphics[width=\linewidth]{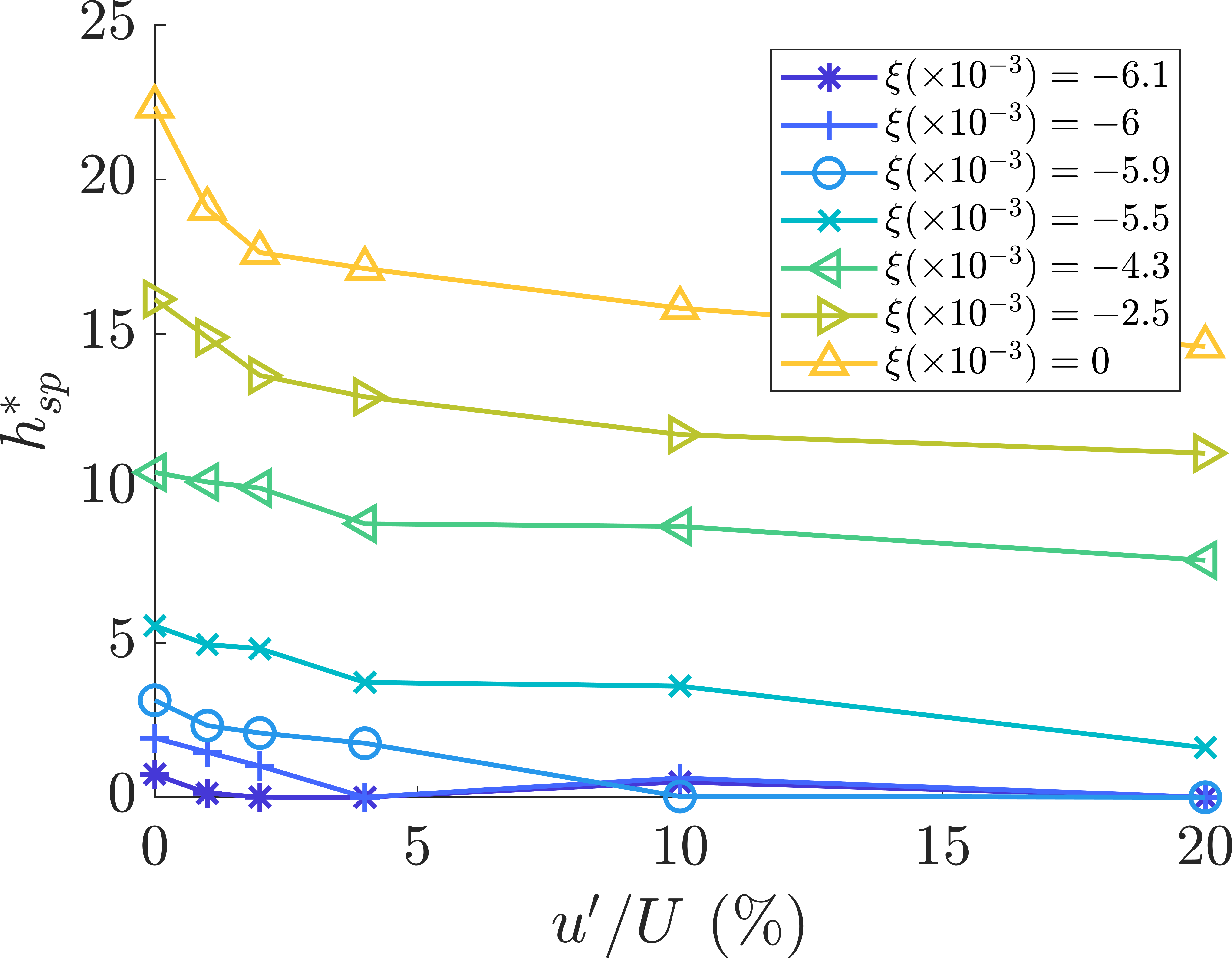}
    \caption{}
        %\caption{Level curves of $h_{sp}^{\ast}$ at different values of $\xi$.}
		\label{fig:hsp_plot}
	\end{subfigure}
\caption{Dimensionless spreading height,
  $h_{sp}^{\ast}$, as a function of its lightness, $\xi$, and
  turbulence level, $u^{\prime}/U$: (a) heatmap; (b) level curves.}
\label{fig:hsp}
\end{figure}

\begin{figure}[htb!]
	\centering
	\begin{subfigure}[b]{0.49\textwidth}
 \includegraphics[width=\linewidth]{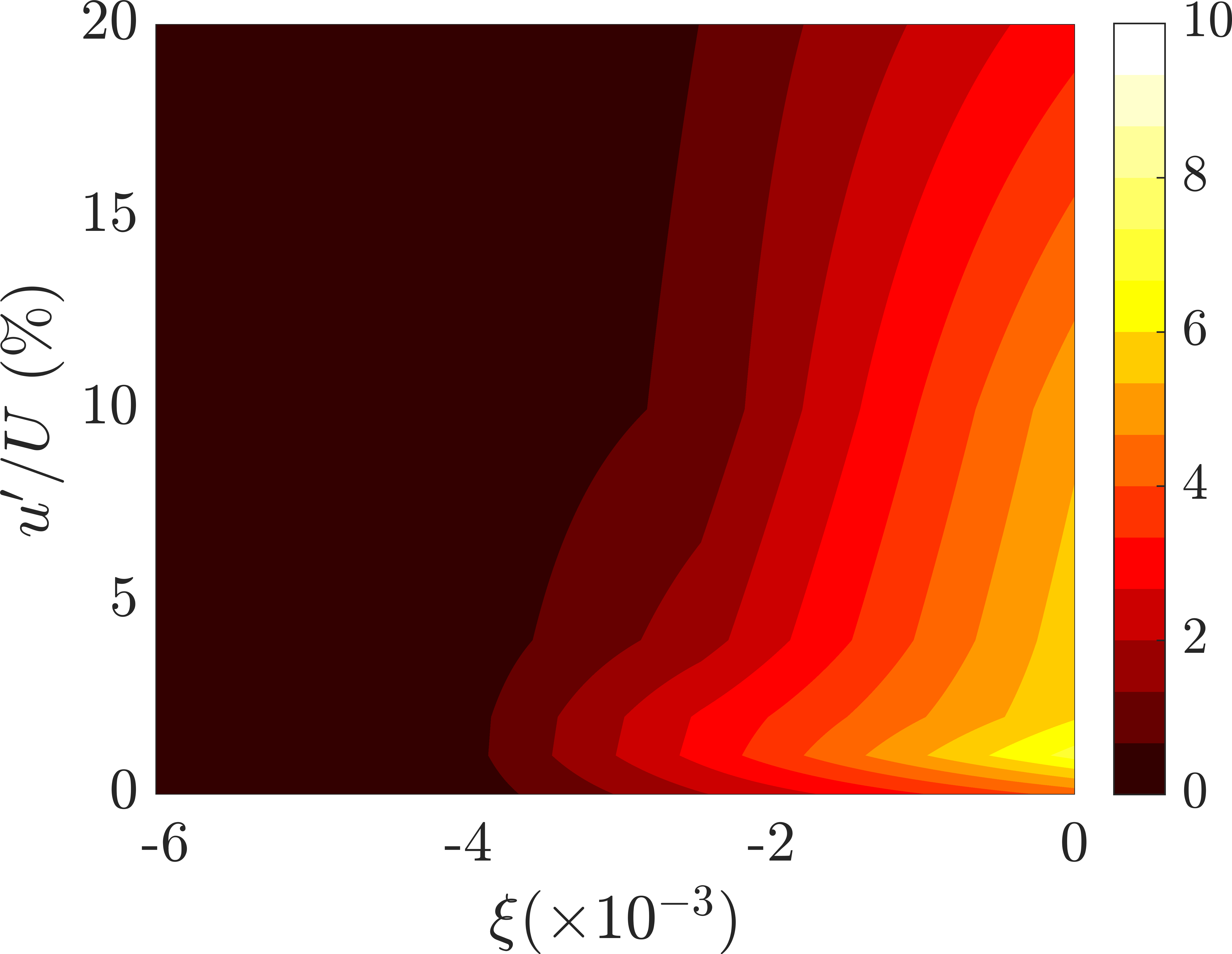}
		\caption{}
        %\caption{Heatmap of $h_{c}^{\ast}(0.01)$.}
		\label{fig:hc1pc_calor}
	\end{subfigure}
	\hfill
	\begin{subfigure}[b]{0.49\textwidth}
 \includegraphics[width=\linewidth]{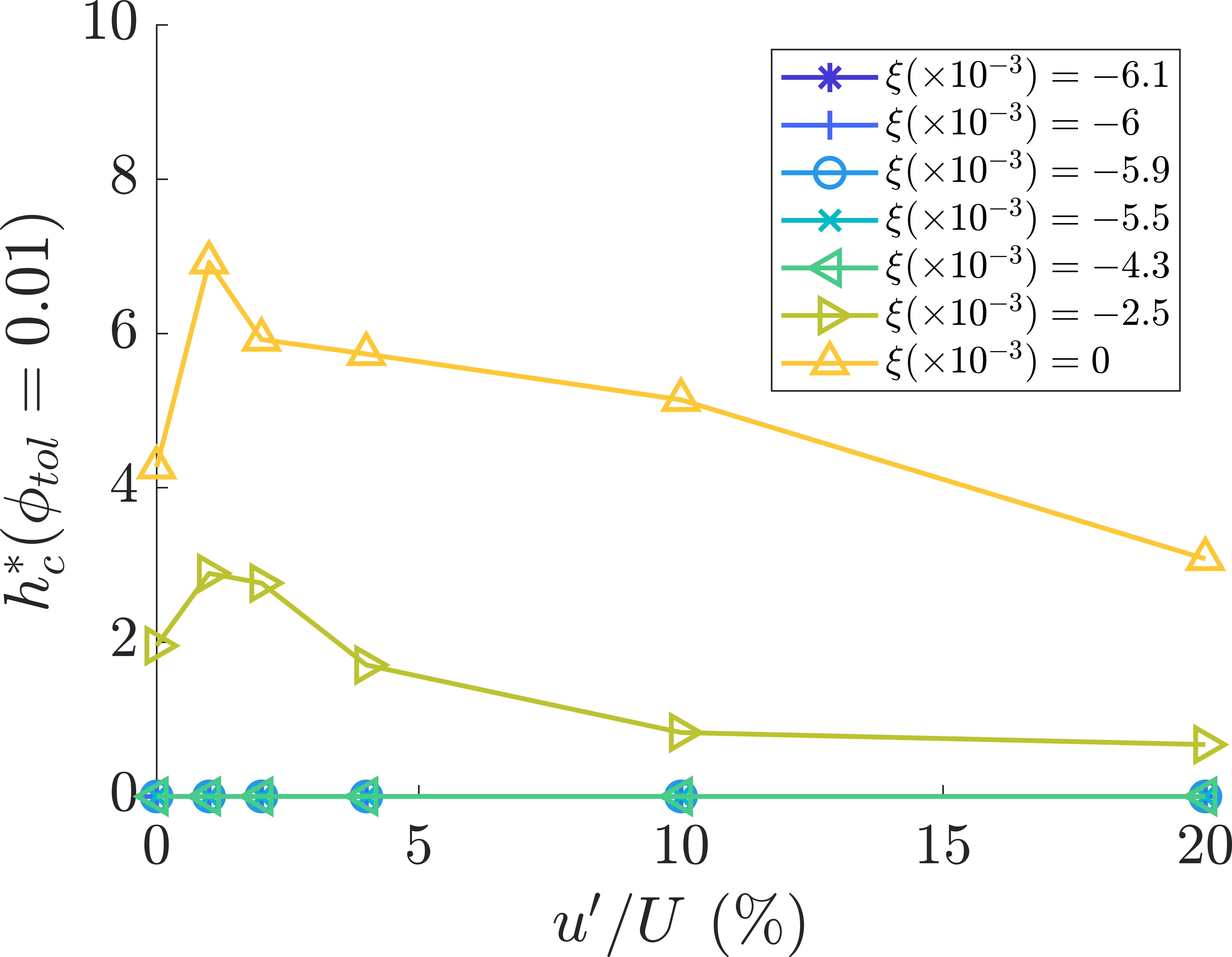}
		\caption{}
        %\caption{Level curves of $h_{c}^{\ast}(0.01)$ at constant $\xi$.}
		\label{fig:hc1pc_plots}
	\end{subfigure}
 \caption{Dimensionless critical height, $h_{c}^{\ast}$, for a tracer
   concentration tolerance of $\phi_{tol}=0.01$: (a) heatmap; (b)
   level curves.}
 \label{fig:hc1pc}
\end{figure}

\begin{figure}[htb!]
	\centering
	\begin{subfigure}[b]{0.49\textwidth}
 \includegraphics[width=\linewidth]{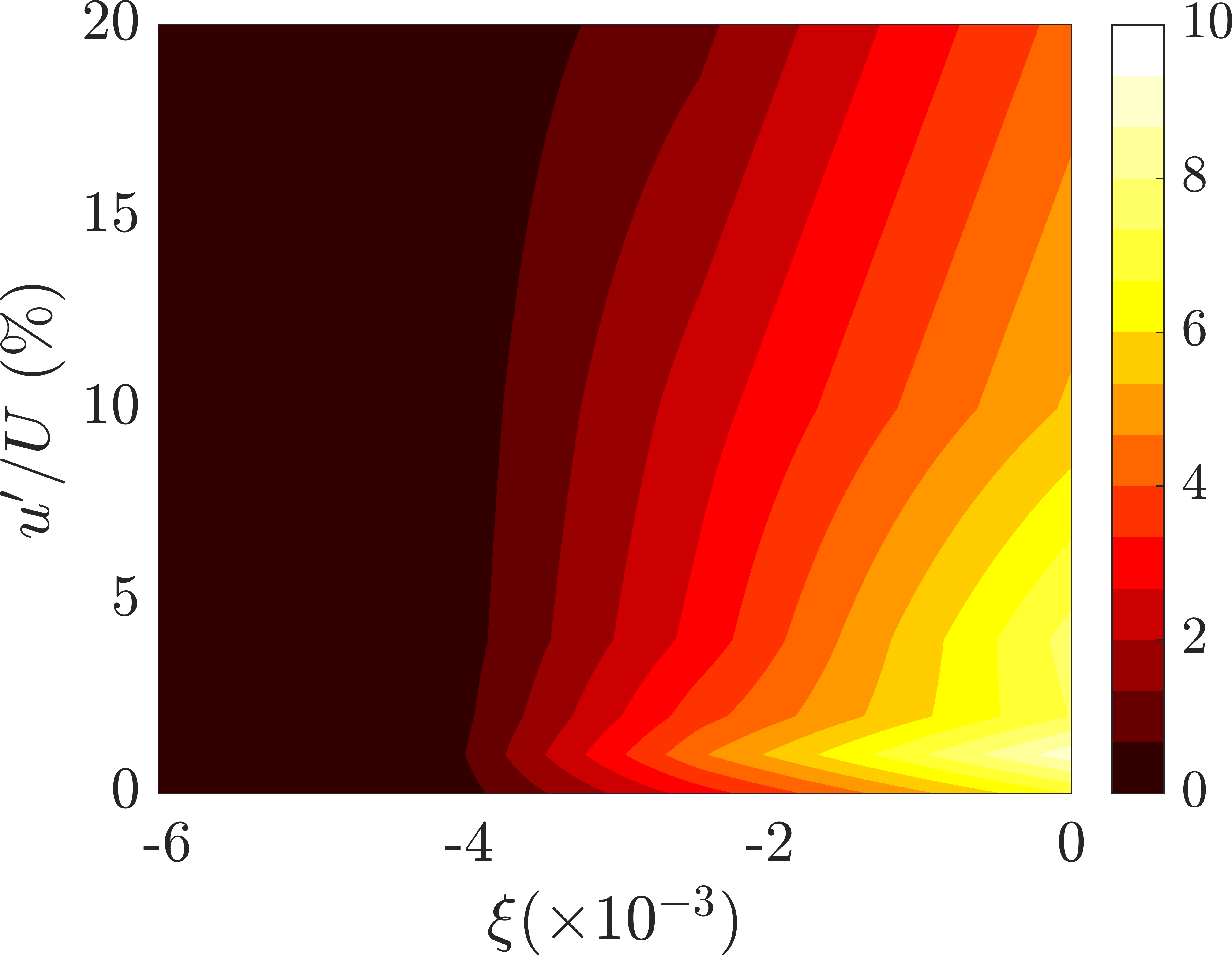}
 \caption{}
        %\caption{Heatmap of $h_{c}^{\ast}(0.05)$.}
		\label{fig:hc5pc_calor}
	\end{subfigure}
	\hfill
	\begin{subfigure}[b]{0.49\textwidth}
 \includegraphics[width=\linewidth]{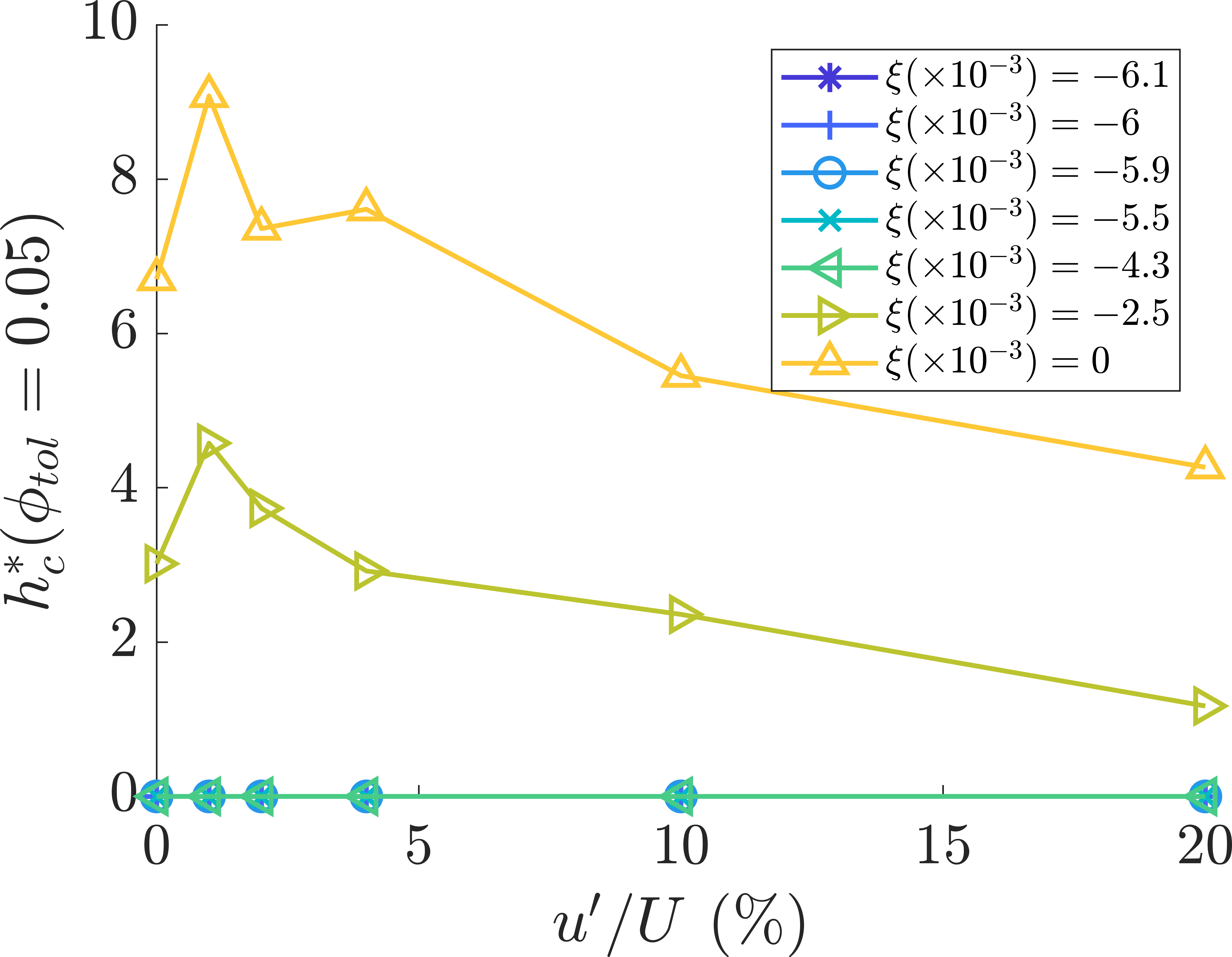}
		\caption{}
        %\caption{Level curves of $h_{c}^{\ast}(0.05)$ at constant $\xi$.}
		\label{fig:hc5pc_plots}
	\end{subfigure}
 \caption{Dimensionless critical height, $h_{c}^{\ast}$, for a tracer
   concentration tolerance of $\phi_{tol}=0.05$: (a) heatmap; (b)
   level curves.}
 \label{fig:hc5pc}
\end{figure}

\begin{figure}[htb!]
	\centering
	\begin{subfigure}[b]{0.49\textwidth}
 \includegraphics[width=\linewidth]{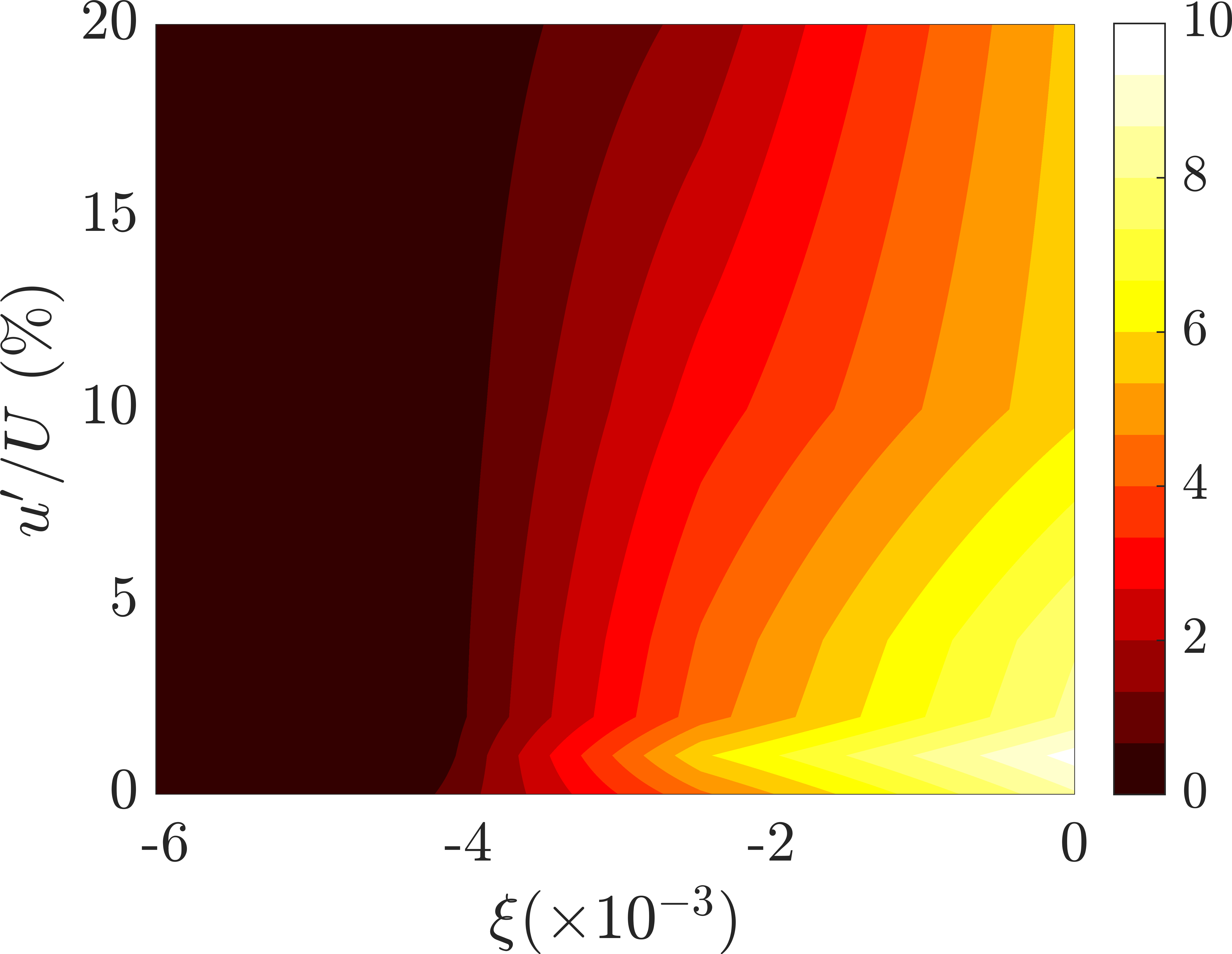}
 \caption{}
        %\caption{Heatmap of $h_{c}^{\ast}(0.20)$.}
		\label{fig:hc20pc_calor}
	\end{subfigure}
	\hfill
	\begin{subfigure}[b]{0.49\textwidth}
 \includegraphics[width=\linewidth]{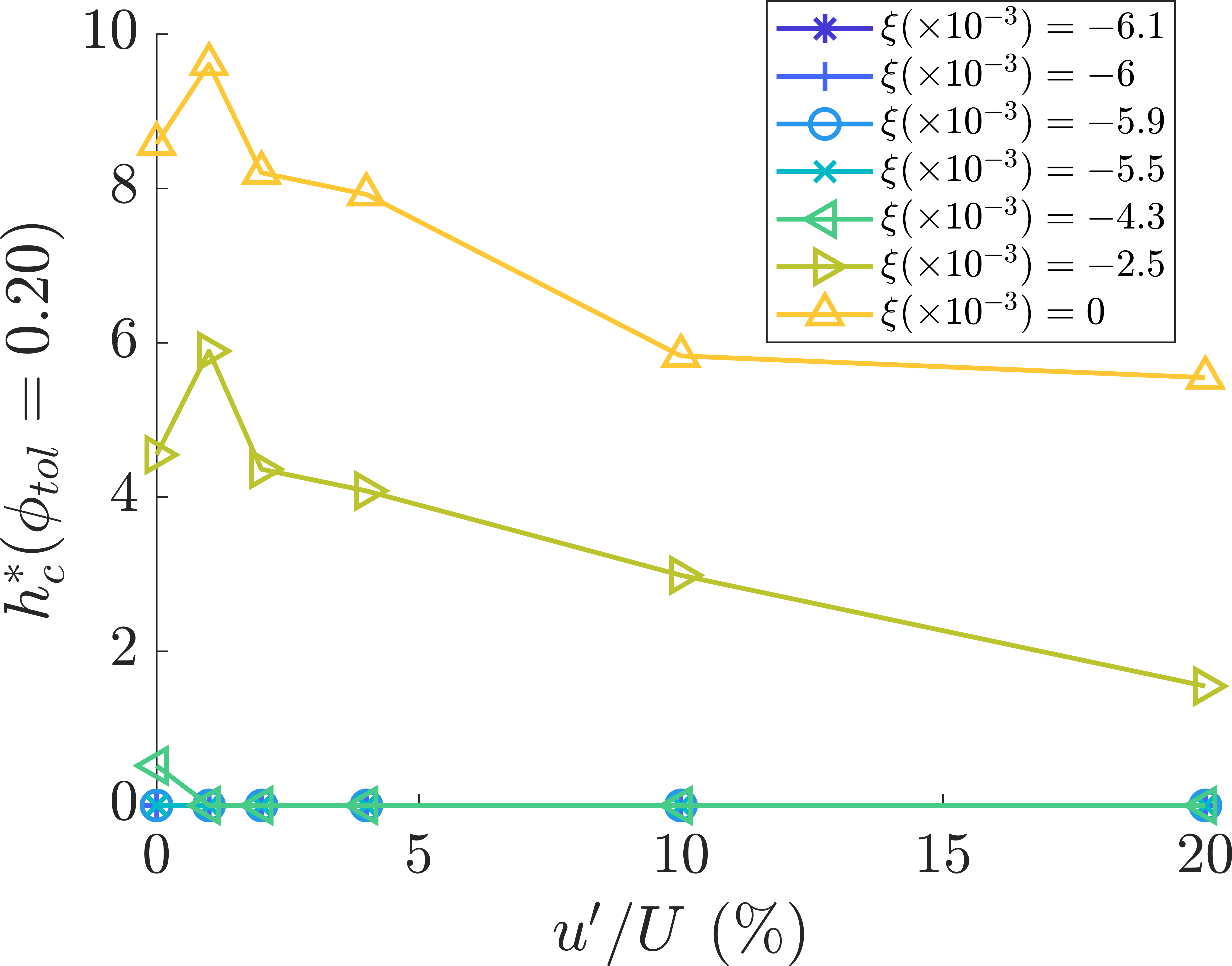}
		\caption{}
        %\caption{Level curves of $h_{c}^{\ast}(0.20)$ at constant $\xi$.}
		\label{fig:hc20pc_plots}
	\end{subfigure}
 \caption{Dimensionless critical height, $h_{c}^{\ast}$, for a tracer concentration tolerance of $\phi_{tol}=0.20$:
 (a) heatmap;  (b) level curves.}
 \label{fig:hc20pc}
\end{figure}

The behavior of the characteristic heights as a function of the
parameter values is crucial to determine the occurrence of the
semi-collapse regime. In Figs.~\ref{fig:hmax} and \ref{fig:hsp} we
show the maximum and spreading heights as a function of the lightness
and turbulence level of the fountain. We observe, in agreement with
previous works~\cite{Sarasua,Freire2010} that, for a given lightness
level, both heights decrease monotonically with the turbulence
level. While for low fountain lightness, the three critical heights
are zero, non-intuitive results are observed for high lightness levels
of the fountain. In such a case, the critical height is not monotonous
but presents a maximum for a certain turbulence level instead. To
check that the results do not depend on the tolerance chosen for the
passive concentration, in Figs.~\ref{fig:hc1pc}, \ref{fig:hc5pc} and
\ref{fig:hc20pc} we show the results for three different values. We
conclude from these figures that the qualitative dependence is very
similar for the three values displayed.


%-------------------------------------
%-------------------------------------

\begin{figure}[htb!]
    \centering
    \includegraphics[width=0.7\linewidth,keepaspectratio]{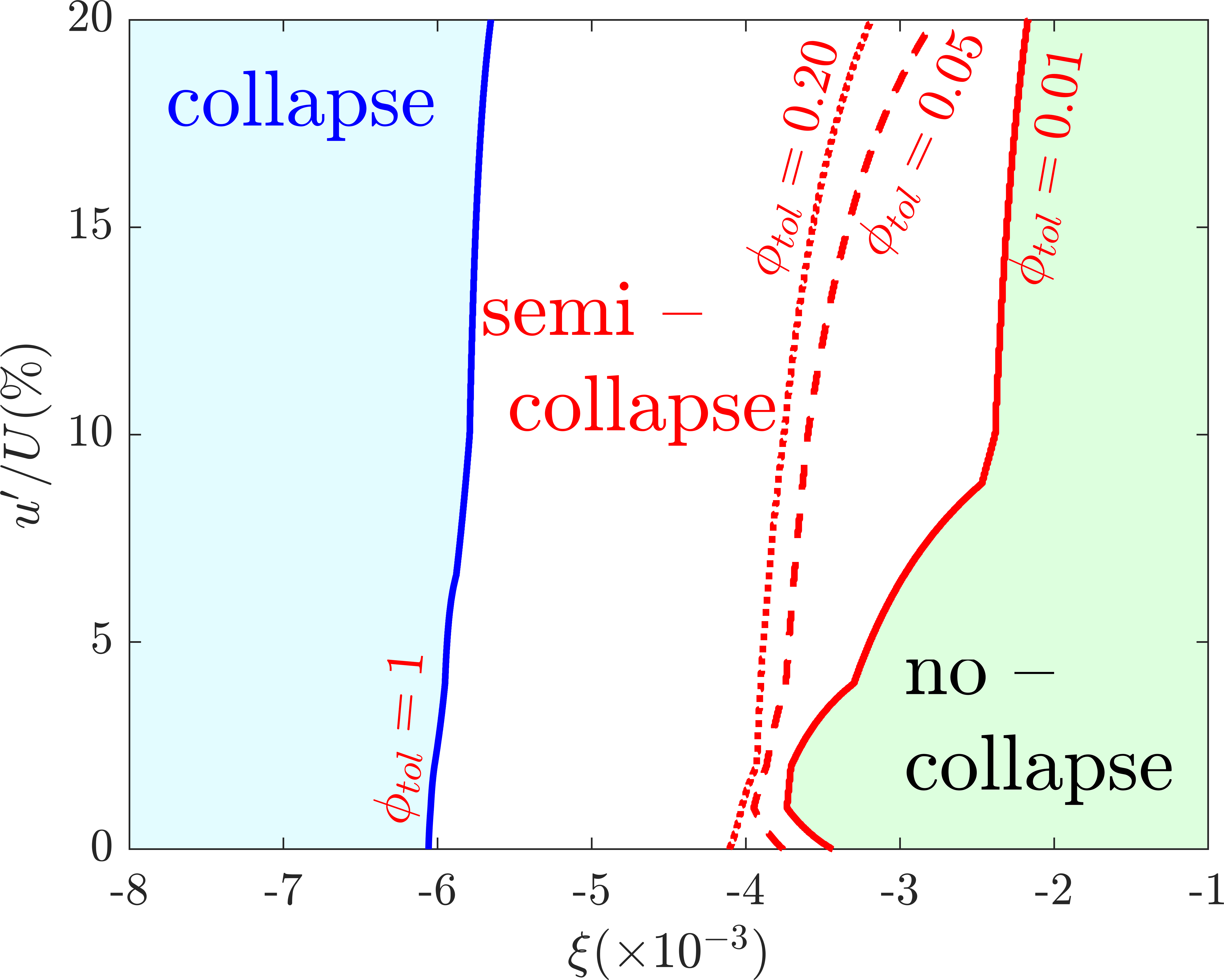}
    \caption{Developed regimes in terms
    %of the fluctuation level $u^{\prime}/U$ and $\xi$ (i.e., in terms of the Froude number).
    of its lightness, $\xi$, and turbulence level, $u^{\prime}/U$:
    collapse (light-blue), semi-collapse (white), and no-collapse
    (green). The red curves delimiting the semi-collapse region depend
    on the concentration thresholds as indicated in the legends.  }
    \label{fig:diag_tinta}
\end{figure}

To summarise the previous results, in the diagram of
Fig.~\ref{fig:diag_tinta}, we condense the results from the previous
graphs.  Such a diagram divides the configuration space lightness --
turbulence level
%\sout{$(\xi, u^{\prime}/U)$}
in three regions, each corresponding to a different behaviours of the
fountain at developed stages: collapse when $h_{sp}\leq0$,
semi-collapse when $h_{sp}>0$ and $h_c>0$ and no-collapse.  Different
boundaries between no-collapse and semi-collapse regions are
determined depending on the value of $\phi_{tol}$.  For non-small
values of $\phi_{tol}$ (meaning a high tolerance to the contaminant),
the critical fountain lightness value for the occurrence of
semi-collapse is weakly dependent on the turbulence level (e.g., line
$\phi_{tol}=0.20$ in Fig.~\ref{fig:diag_tinta}).  However, for small
values of $\phi_{tol}$ (e.g., $\phi_{tol}=0.01$), such a boundary
boundary is not monotonous.  It can be observed from the diagram that
the desirable non-collapse regime is more stable at a certain low
fluctuation level since the no-collapse region is broader there.  In
addition, from this figure, it is observed the no-collapse region
boundary for $\phi_{tol}=0.01$ presents a minimum with respect to the
turbulence level.
%$u^{\prime}/U$.
From Fig.~\ref{fig:diag_tinta}, the such minimum is located about the coordinates
%\sout{$p=$}
$(\xi=-3.7\times10^{-3}, u^{\prime}/U=1\%)$.  For technical
applications of fountains for the removal of contaminants, as
explained in Sec.~\ref{sec:terreno}, such behaviour means that
starting from semi-collapse configurations whose lightness level is
$\xi>-3.7\times10^{-3}$, transition to the desired no-collapse regime
can be induced by decreasing the turbulence level of the fountain if
$u^{\prime}/U>1\%$, or increasing such turbulence level if
$u^{\prime}/U<1\%$. The latter is a surprising result and is a key
contribution of the present work, not reported before to the best of
the authors' knowledge.

Meanwhile, regarding the lightness level of the fountain for a given
turbulence level, the transition from a semi-collapse to a no-collapse
regime can only be done by increasing the fountain's lightness, i.e.,
increasing the fountain temperature $T_{in}$.

%-------------------------------------

\subsection{Effectiveness on the contaminant removal}
\label{sec:terreno}

In addition to this primary application, the SIS device offers various
other notable features and applications~\cite{sis_technologies}. The
device operates at a lightness level of $\xi=0$ by selectively
removing the lower strata of air under thermal inversion
stratification conditions. The average diameter of the SIS is
approximately $3-4$~m and it is designed to protect the surrounding
area by redirecting the removed air above a critical height of
$h_c\geq 6-8$m, which corresponds to a height of at least
$2D$\cite{sis_technologies}. From Figs.~\ref{fig:hc1pc} to
\ref{fig:hc20pc}), we corroborate that it is achieved only for
$\xi=0$.  In fact, during operation, the spreading height in the field
is about 50~m, which is in excellent agreement with our results.
%From Figs.~\ref{fig:hc1pc} to \ref{fig:hc20pc}), we concluded that it is achieved only for $\xi=0$.
Moreover, from Fig.~\ref{fig:hc1pc}, it is also feasible for
$\xi=-2.5\times10^{-3}$ and $u^{\prime}/U\leq 2\%$, meaning that if
the fountain is colder than the air close to the ground, the air would
still be clean below $2D$ if the turbulence intensity is small.

\begin{figure}[htb!]
	\centering
	\begin{subfigure}[b]{0.495\textwidth}
  \includegraphics[width=\linewidth]{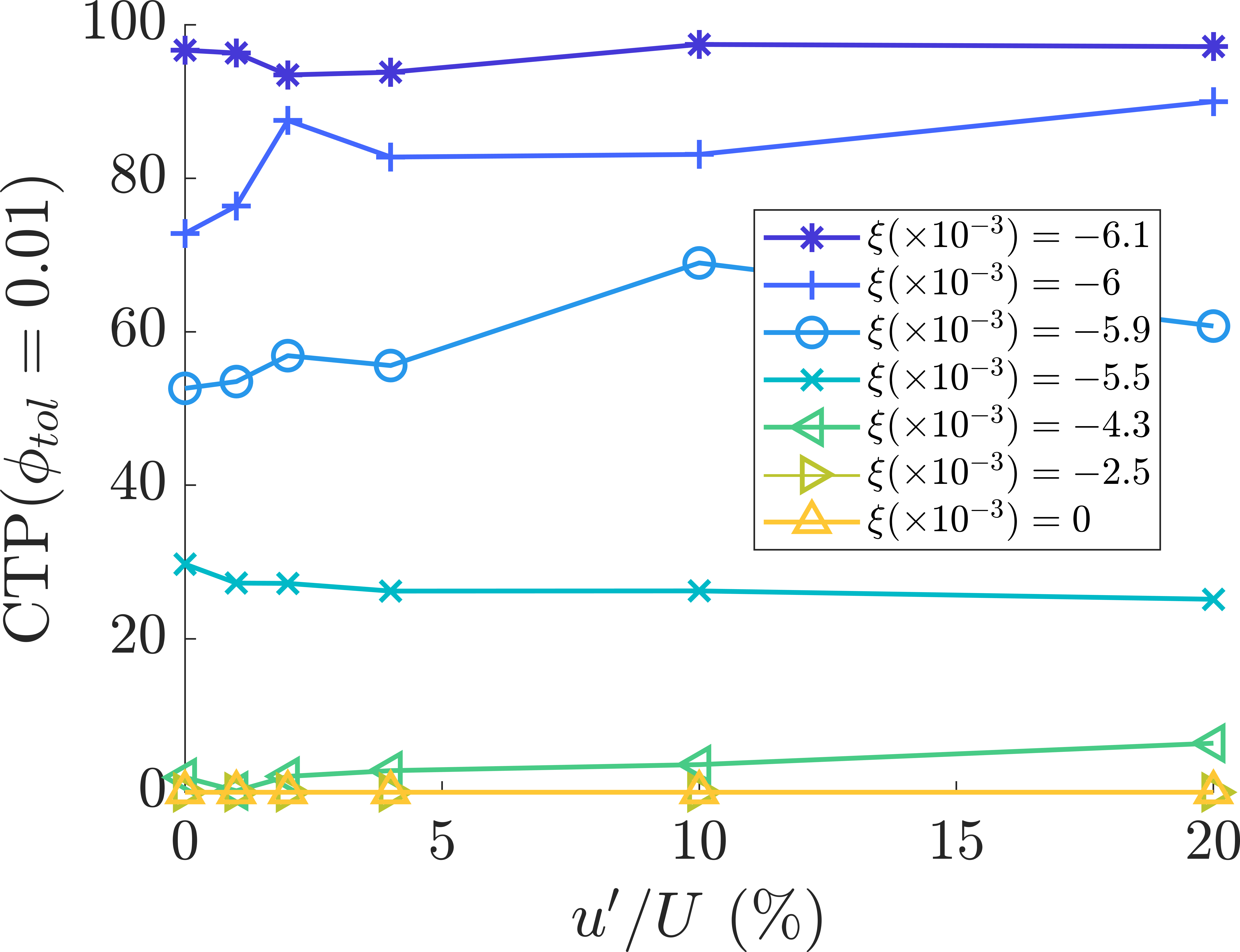}
		\caption{}
		\label{fig:ctp1pc}
	\end{subfigure}
	\hfill
	\begin{subfigure}[b]{0.495\textwidth}
     \includegraphics[width=\linewidth]{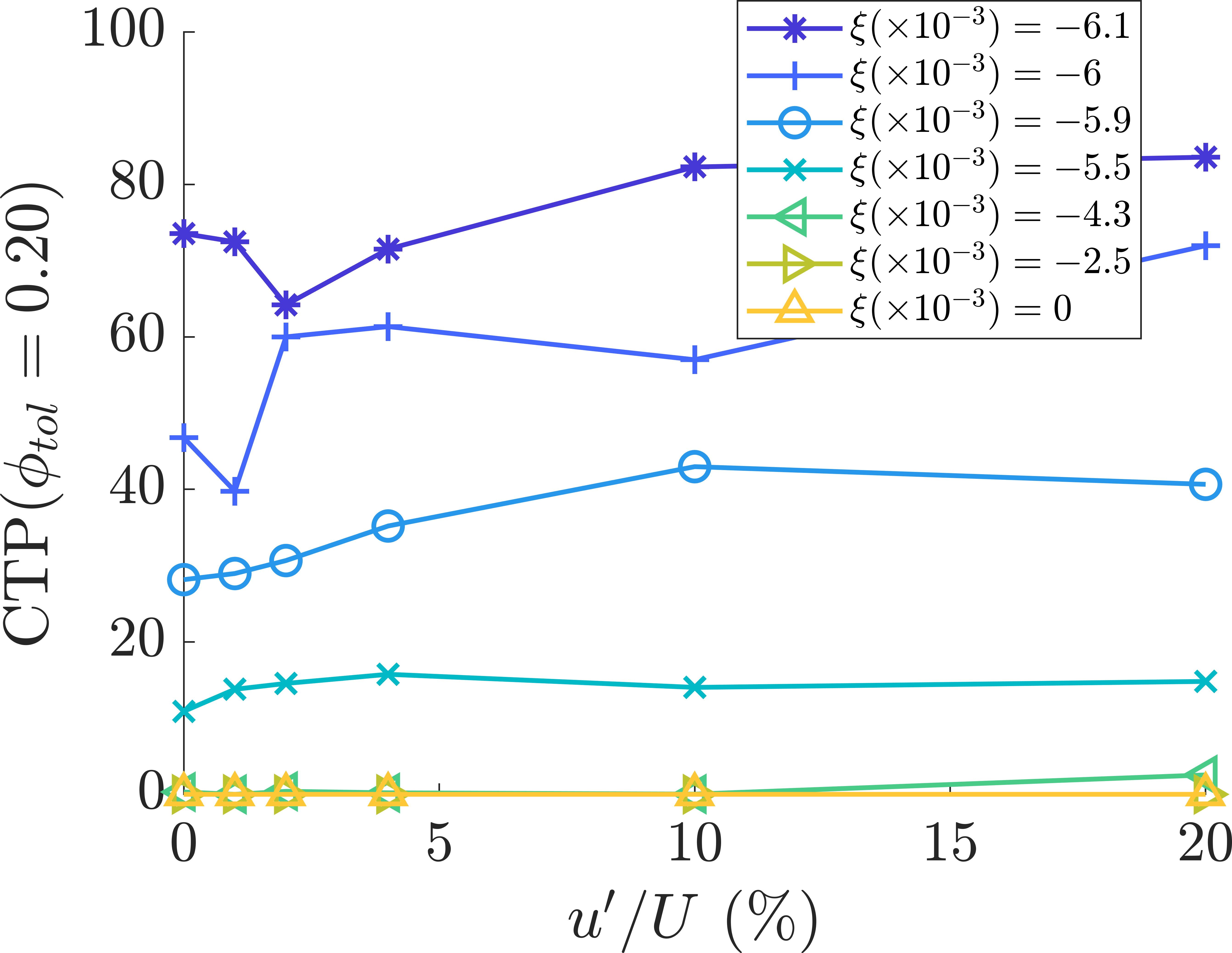}
    \caption{}
        %\caption{CTP for $\phi_{tol}=0.20$.}
    \label{fig:ctp20pc}
    \end{subfigure}
	\caption{Contaminated terrain percentage (CTP) as a function
          of the turbulence level, $u^{\prime}/U$, for different
          values of the lightness $\xi$. Tolerance values of the
          passive scalar concentration: (a)
          $\phi_{tol}=0.01$ and (b) $\phi_{tol}=0.20$.}
        \label{fig:ctp}
\end{figure}

A key aspect is the percentage of the terrain that is affected by
contaminants under semi-collapse conditions. For a given value of
$\phi_{tol}$, we define the contaminated terrain percentage,
CTP($\phi_{tol}$), as the total area of the plane at the ground
($z=0$) where $\phi\geq\phi_{tol}$ divided by the total area.  The
results of CTP for $\phi_{tol}=0.01$ and $0.20$ are shown in
Fig.~\ref{fig:ctp1pc} and \ref{fig:ctp20pc}, respectively.  In case of
not so strict requirements, for instance, CTP$\leq 10\%$, from
Fig.~\ref{fig:ctp}, we conclude that every configuration with $\xi\geq
-4.3\times10^{-3}$ (i.e., $T_{in}\geq 10 \ ^{\circ}$C) is safe, no
matter the level of turbulence.

\section{Final remarks} \label{sec:conc}
In this study, we investigated the collapse of turbulent fountains,
focusing on the impact of the turbulence level and a characteristic
parameter known as lightness. This parameter is defined by the
temperature difference between the fountain and the surrounding
ambient fluid at the fountain inlet (ground) level.  We identified
three distinct regimes based on the spreading height,$h_{sp}$, and the
minimum or \textit{critical} height, $h_{c}$: collapse regime for
$h_{sp}\leq0$; semi-collapse regime for $h_{sp}>0$ and $h_c>0$; and
no-collapse regime for $h_{sp}>0$. To effectively monitor and measure
the characteristic heights of the flow, we introduced a passive scalar
tracer concentration field into the inflow. The determination
of the critical height $h_c$ is a delicate process that relies heavily
on the selected tracer tolerance level.

Our results are consistent with previous research indicating that an
increase in turbulence level reduces both $h_{m}$ and
$h_{sp}$. However, the dependence of $h_c$ on the turbulence level is
non-monotonic and strongly influenced by the $\phi_{tol}$. For large
enough values of $\phi_{tol}$, the critical lightness value for the
occurrence of semi-collapse is weakly dependent on the turbulence
level. Remarkable, for lower values of $\phi_{tol}$, such as
$\phi_{tol}=0.01$, the critical height $h_c$ exhibits a non-monotonic
behavior. As a result, the no-collapse regime is more stable at a
certain low fluctuation level, as shown in
Fig.~\ref{fig:diag_tinta}. This allows for a transition from the
semi-collapse regime to the no-collapse regime by increasing the
turbulence level.  Additionally, two possible transitions from a
semi-collapse to a no-collapse configuration are possible under
certain conditions, either by decreasing or increasing the turbulence
level. However, to observe such a transition the fountain lightness
must be increased.

Finally, we applied our analysis to evaluate the effectiveness of
turbulent fountains in removing pollutants modelled by a the scalar
field concentration. Of particular interest was the semi-collapse
scenario, where we examined the percentage of contaminated terrain
where $\phi\geq\phi_{tol}$. Our findings revealed that regardless of
$\phi_{tol}$, the proportion of contaminated terrain remained below
$10\%$ for $\xi\geq-4.3\times10^{-3}$ (i.e., $T_{in}\geq10.0
\ ^{\circ}$C), independent of the turbulence level. Thus, our analysis
provides a powerful tool for enhancing the efficiency of technological
applications, such as the SIS device.

%\section*{Acknowledgements} \label{sec:ack}
%The authors would like to thank PEDECIBA (MEC, UdelaR, Uruguay) and
%express their gratitude for the grant Fisica Nolineal (ID 722)
%Programa Grupos I+D CSIC 2018 (UdelaR, Uruguay).

\section*{\label{sec:data} Data availability}

The data that support the findings of this study are available from
the corresponding author upon reasonable request.

\section*{\label{sec:decla} Declarations}

\noindent \textbf{Competing interests:} The authors have no competing
interests or other interests that might be perceived to influence the
results and/or discussion reported in this paper.

\noindent \textbf{Authors' contributions:} Luis G. Sarasúa conceived
the numerical experiment, analysed and discussed the results, and
contributed to manuscript writing. Daniel Freire Caporale performed
the laboratory and numerical experiments, analysed the results,
prepared the figures, and contributed to manuscript writing. Nicasio
Barrere contributed to the numerical experiment and analysis of the
results. Arturo C. Martí participated in the experiment's design,
contributed to the analysis of the results, prepared the final
version, and reviewed the manuscript.

\noindent \textbf{Funding:} The received partial fuding by PEDECIBA (MEC, UdelaR, Uruguay) and
 grant Fisica Nolineal (ID 722) Programa Grupos I+D CSIC 2018 (UdelaR, Uruguay).

%\bibliography{biblio_semi_colapso}% common bib file



\end{document}